\documentclass[pre,onecolumn,11pt,tightenlines,amsmath,amssymb,superscriptaddress,runinaddress,letterpaper,floatfix,nofootinbib]{revtex4}

\usepackage{upgreek} 
\usepackage{graphicx} 
\usepackage{amsmath}
\usepackage{txfonts}
\usepackage{amssymb}
\usepackage{times}
\usepackage{booktabs} 

\begin{document}

\title{Mechanical fluidity of fully suspended biological cells
}

\author{John M.\ Maloney}
\affiliation{%
Department of Materials Science and Engineering \\
Massachusetts Institute of Technology, Cambridge, MA 02139 USA
}%
\author{Eric Lehnhardt}
\affiliation{%
School of Biological and Health Systems Engineering \\ 
Arizona State University, Tempe AZ USA\\
}%
\author{Alexandra F.\ Long}
\affiliation{%
Department of Biology \\ 
Carleton College, Northfield MN USA}%
\author{Krystyn J.\ Van Vliet}%
\thanks{Address: Laboratory for Material Chemomechanics (8-237); Massachusetts Institute of Technology; 77 Massachusetts Avenue; Cambridge, MA 02139}
\email{krystyn@mit.edu}
\affiliation{%
Department of Materials Science and Engineering and Department of Biological Engineering\\
Massachusetts Institute of Technology, Cambridge, MA 02139 USA
}%

\begin{abstract}
Mechanical characteristics of single biological cells are used to identify and possibly leverage interesting differences among cells or cell populations. Fluidity---hysteresivity normalized to the extremes of an elastic solid or a viscous liquid---can be extracted from, and compared among, multiple rheological measurements of cells: creep compliance vs.\ time, complex modulus vs.\ frequency, and phase lag vs.\ frequency. With multiple strategies available for acquisition of this nondimensional property, fluidity may serve as a useful and robust parameter for distinguishing cell populations, and for understanding the physical origins of deformability in soft matter. Here, for three disparate eukaryotic cell types deformed in the suspended state via optical stretching, we examine the dependence of fluidity on chemical and environmental influences around a time scale of 1\,s. We find that fluidity estimates are consistent in the time and the frequency domains under a structural damping (power-law or fractional derivative) model, but not under an equivalent-complexity lumped-component (spring-dashpot) model; the latter predicts spurious time constants. Although fluidity is suppressed by chemical crosslinking, we find that adenosine triphosphate (ATP) depletion in the cell does not measurably alter the parameter, and thus conclude that active ATP-driven events are not a crucial enabler of fluidity during linear viscoelastic deformation of a suspended cell. Finally, by using the capacity of optical stretching to produce near-instantaneous increases in cell temperature, we establish that fluidity increases with temperature---now measured in a fully suspended, sortable cell without the complicating factor of cell-substratum adhesion. 
\end{abstract}

\maketitle

\section{Introduction}
Biological tissue cells are arguably the preeminent mechanical material to be understood---no other material is so complex while so intimate to our existence. The capacity to parameterize the mechanical response of such cells to applied loads informs our understanding and modeling of structurally dynamic, contractile polymer networks. Further, a distinct mechanical signature can potentially enable the sorting of useful or diseased cells from mixed populations. To this end, researchers have quantified the rheology (deformation and flow characteristics) of single, animate cells~\cite{fabry2001scaling,balland2006power,hoffman2006consensus,chowdhury2008cell}  and of inanimate soft condensed matter comprising cytoskeletal and motor proteins~\cite{gardel2006stress}. Such studies have included analysis of both internal~\cite{tseng2002micromechanical,wilhelm2008out,robert2012magnetic} and cortical~\cite{stamenovic2002effect,alcaraz2003microrheology,lenormand2004linearity,hiratsuka2009power,kollmannsberger2010nonlinear} deformability of attached and contractile cells. Others have also explored chemical modulation of metabolism and cytoskeletal rearrangements~\cite{bursac2005cytoskeletal,trepat2008universality}  to elucidate molecular origins of single-cell stiffness and contraction. Although fewer studies have considered the rheology of cells in the nominally detached or fluid-suspended state~\cite{roca2006rheology,maloney2010,macqueen2010mechanical},  this state is more relevant to practical applications of cell biophysics to technological and medical applications. For example, identification and isolation of valuable cells from mixed populations (e.g., circulating tumor cells or stem cells) may rely wholly or in part on mechanical signatures of cells dispersed in solution~\cite{guck2005optical,hur2011deformability,adamo2012microfluidics,zhang2012microfluidics,preira2013passive}. Given the potential for comparatively higher throughput analysis of such cells in the suspended state, it is reasonable to expect that biophysical characterization of whole, suspended cells will continue to inform diagnostic assays~\cite{guck2005optical}, injections of cells for targeted delivery~\cite{lunde2006intracoronary}, and basic understanding of tissue cells that lack cytoskeletal stress fibers when located within highly compliant, three-dimensional tissues or synthetic constructs~\cite{byers1984organization,ong2008gel,khatau2012distinct}.

To evaluate biophysical models or to compare cells (or cell populations) quantitatively, mechanical behavior is often parameterized by the complex modulus, which reports both the stiffness and viscoelastic damping or hysteresivity.  
Here, we focus on a single parameter---fluidity $a$, or normalized hysteresivity---that is related to the position of the cell in a solid-liquid continuum of soft matter.\footnote{It is possible to calculate this parameter from, for example, the phase lag of sinusoidal deformation caused by sinusoidal loading in the linear regime. One can consider this phase lag to be bounded by zero (corresponding to an elastic solid) and  $\pi/2$ radians or one quarter period (corresponding to a viscous liquid); normalizing to these extremes produces a fluidity value~\cite{pajerowski2007physical,klemm2009comparing,coughlin2012cytoskeletal} that measures the tendency of the cell to flow (vs.\ stretch and rebound) in response to a mechanical load.}
As here we seek to identify potential differences among single-cell mechanical parameters (as a function of cell type and chemical and physical environment), the length scale of interest in extracting and discussing fluidity is the whole cell, while the relevant time scale for the current work is $\sim$1\,s. 
At that length scale, the cell is considered as a viscoelastic material that is also a spatially heterogeneous composite comprising an actin cortex, cytoplasm, nucleus and numerous organelles~\cite{hoffman2006consensus,hoffman2009cell}. We note that, at this time scale and related frequency, high-throughput sorting of individual cells is plausible. However, it is known that other time scales exhibit distinct features; specifically, at much higher frequencies, the contribution of water viscosity predominates~\cite{fabry2001scaling,moeendarbary2013cytoplasm}, while at much longer times cytoskeletal rearrangements and remodeling in response to loads become measurable~\cite{icard2008cell}. 

Importantly, the parameter referred to here as fluidity is independent of models used to interpret and predict cell deformation. However, fluidity measurements can be used to evaluate models of soft matter that are applied to whole-cell deformation around our time scale of interest. For example, deformation behavior has often been modeled with an assembly of several springs and dashpots~\cite{feneberg2004microviscoelasticity,wottawah2005optical,wottawah2005characterizing,lu2006viscoelastic,rosenbluth2008slow,teo2010cellular,nguyen2010biomechanical}. Here, models predict time constants~\cite{wottawah2005optical,wottawah2005characterizing} near 1\,s, corresponding presumably to cytoskeletal biophysical mechanisms, at which relatively large changes in fluidity are mathematically predicted. Alternatively, others have used a structural damping or fractional derivative model in which  creep compliance vs.\ time, complex modulus vs.\ frequency, and stress relaxation vs.\ time all appear as power laws~\cite{fabry2001scaling,fabry2003time,roca2006rheology,massiera2007mechanics,hemmer2009role,hiratsuka2009power,zhou-power}. Here, fluidity---equivalent to the power-law exponent---is viewed as frequency independent~\cite{alcaraz2003microrheology,balland2006power} at time scales near 1\,s. Even within the neighborhood of our time scale of interest, therefore, the question remains of which model is best suited to parameterize whole cells accurately; especially considering the smaller number of studies of  cells in the suspended state, uncertainty also exists surrounding whether contractile stress fibers are needed for the cell to manifest power-law rheology~\cite{kollmannsberger2009active}. 

\begin{figure*}[t]
\centering
\includegraphics{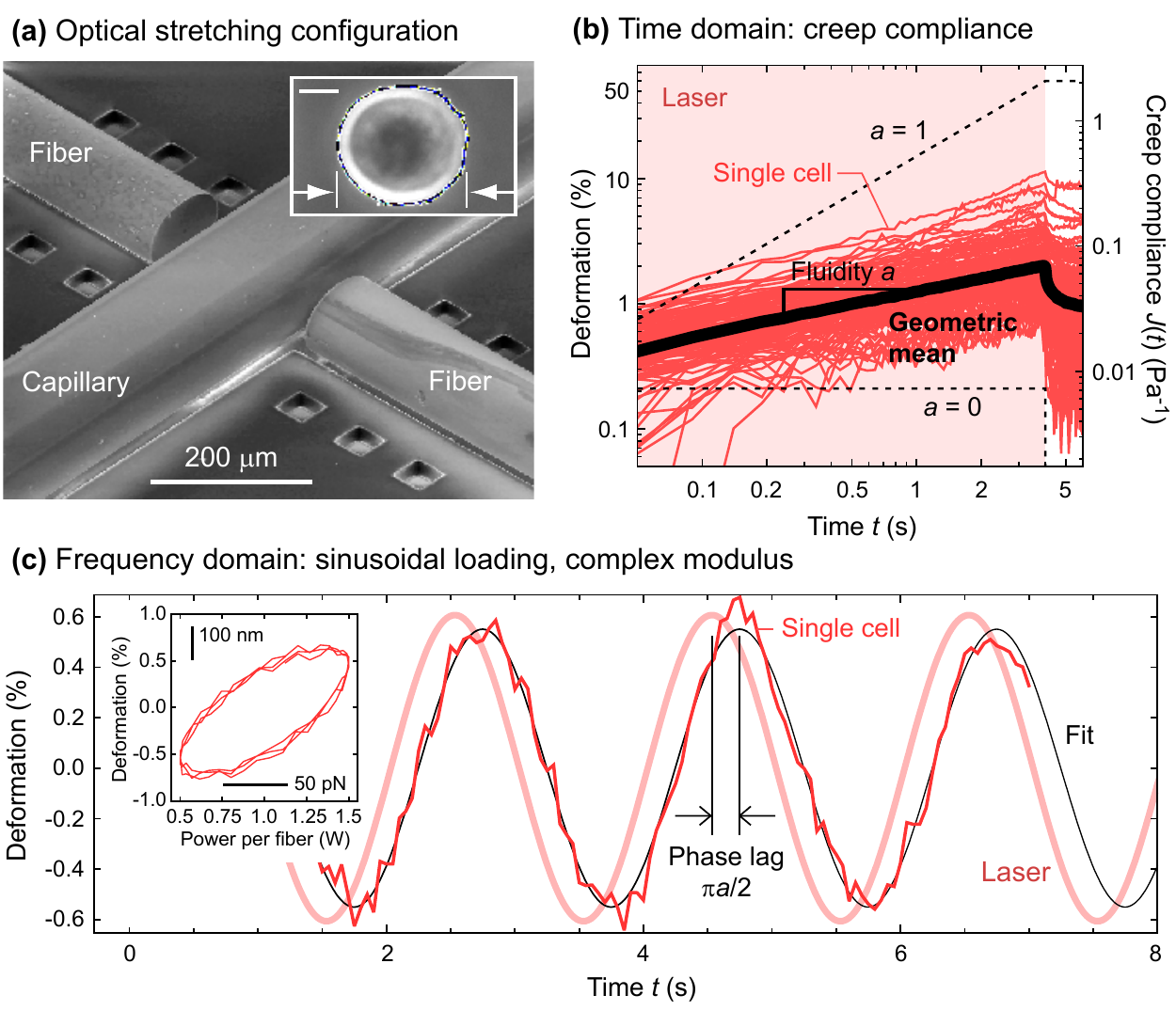}
\caption{Optical stretching (OS) measures the stiffness of cells in the suspended state, absent physical contact and direct influence of substratum chemomechanical properties. (a)~Scanning electron micrograph of opposing optical fibers  positioned to direct laser emission toward a hollow glass capillary filled with cell suspension. (During operation, fibers are surrounded in index-matching gel and positioned approximately 100\,$\upmu$m away from the capillary wall.) Inset, edge detection applied to a phase contrast image to quantify cell deformation upon photonic loading; scale bar = 10\,$\upmu$m. (b)~From one eukaryotic cell type (CH27 lymphoma, $n=121$ cells),  thin red lines show  deformation and creep compliance $J(t)$ for single cells in response to a step increase in laser power from time $t=0$\,s to 4\,s. Thick black line shows  geometric mean, well fit during stretching by the relationship $J(t)\propto t^a$ where $a$ is a measure of cell fluidity. Dotted black lines contrast behavior of perfectly elastic ($a=0$) and viscous ($a=1$) materials in creep compliance stretching and recovery (vertical positioning of these lines is arbitrary). (c)~Oscillatory deformation (minus baseline, see Supp.\ Fig.~S\ref{fig:sine-fitting}) 
 of a single cell in response to sinusoidal loading with angular frequency $\omega$. (Inset, symmetric and elliptical Lissajous figure indicates linear viscoelasticity.) The  viscoelastic phase lag $\phi$ of the cell in radians is also a measure of cell fluidity as $a=2\phi/\pi$; thus, fluidity can be estimated through experiments  in both the time and frequency domains.}
\label{fig:OS}
\end{figure*}

We investigate and quantify the fluidity of the suspended cell via optical stretching, a technique requiring no cell-probe or cell-substratum contact. In optical stretching, dual counterpropagating laser beams attract and center a single suspended cell, which deforms by outward photon-induced stress caused primarily by the change in refractive index at the cell edge (Fig.~\ref{fig:OS})~\cite{guck2001thesis,guck2001optical}. The cell response is typically characterized by its deformation along the laser axis as a function of time. This approach enables us to elucidate how suspended cells deform by removing the influence of stress fibers and adhesion sites, by probing cells in both the time and frequency domains, and by testing cells exposed to chemical and physical perturbation. We consider three distinct, model cell types: (1) the adult human bone-marrow-derived mesenchymal stem or stromal cell (hMSC), which undergoes the attached-to-suspended transition repeatedly during passaging, most notably at the last detachment immediately before re-implantation for therapeutic purposes; (2) the transformed---or immortalized---murine fibroblast (3T3), which is relatively easy to culture due to rapid proliferation and is commonly used as a model cell in rheology studies~\cite{hiratsuka2009power,kollmannsberger2010nonlinear,zhou-power}; and (3) the transformed and nonadherent murine lymphoma cell (CH27), which exhibits no substratum attachment response and thus does not exhibit contractile stress fibers prior to optical stretching. Testing in both the time and frequency domains allows estimation of the fluidity by three methods---the slope of the creep compliance response vs.\ time on a log-log scale,  the phase lag under an oscillatory load, and the slope of the complex modulus response vs.\ frequency on a log-log scale---to reduce the role of chance and experimental artifacts when testing the predictions of various viscoelastic models. We find that fluidity is frequency independent for multiple  cell types and also upon depletion of adenosine triphosphate (ATP), and is further an increasing function of cell temperature. Consideration of this nondimensional rheological characteristic of single cells can thus enable rapid measurements and new predictions relevant to mechanisms of single-cell deformation; obtaining this parameter in the suspended state furthers understanding relevant to physical sorting and delivery of suspended cells. 

\section{Materials and Methods}

\subsection{Cell culture} Primary adult human mesenchymal stem cells (hMSCs) were isolated from the bone marrow of four adult donors (via Stem Cell Technologies, Inc.\, ReachBio, Inc., or Lonza Group Ltd.), cultured in proprietary media (basal with 10\% supplements, Stem Cell Technologies \#5401 and \#5402), and used between passages 1 and 9, as described previously~\cite{maloney2010}. Transformed murine NIH 3T3 fibroblasts were obtained from ATCC (\#CRL-1658) and cultured in DMEM (Gibco \#11885) with 10\% fetal bovine serum (Atlanta Biologicals \#S11550). Immortalized murine CH27 lymphoma cells~\cite{haughton1986ch} were obtained courtesy of  D.~J.~Irvine (MIT) and cultured in RPMI (Gibco \#11875) with 10\% fetal bovine serum (Atlanta Biologicals \#S11550). 

Chemical fixation  was accomplished by incubating  suspended cells at a concentration of approximately 100K\,cells\,mL$^{-1}$ in a 25\% glutaraldehyde-water solution diluted in phosphate-buffered saline (PBS) complete media  for 10 min  at 37$^\circ$C. The suspensions were then diluted $50\times$, centrifuged, and resuspended in PBS for optical stretching. 

Adenosine triphosphate (ATP) was depleted by exposing 3T3 fibroblasts to the standard cocktail of 0.05\% sodium azide  and 50\,mM 2-deoxyglucose~\cite{bursac2005cytoskeletal,hoffman2006consensus}. The degree of ATP depletion was assayed by  luciferase assay and was found to be \{95\%, 96\%, 96\%, 98\%\} in four replicate experiments. Preliminary ATP depletion experiments were performed both before and after trypsinization of 3T3 fibroblasts. When performed before trypsinization, the suspended cells were not spherical (Fig.~\ref{fig:chemicals}(b(i,ii))), indicating that cell remodeling processes initiated by trypsinization and detachment could not be completed, and confirming that active cytoskeletal processes were interrupted by ATP depletion. All stretching experiments were thus performed by depleting ATP after the cells were detached and allowed to remodel in the suspended state at 37$^\circ$C for 1 hour, which is sufficient for remodeling processes to complete~\cite{maloney2010} and which resulted in near-spherical cells in the detached state. 

\subsection{Microfluidic optical stretching and data analysis} Optical stretching and subsequent data analysis in the time domain were conducted generally as described previously~\cite{guck2001optical,lincoln2007reconfigurable,maloney2010}, with differences noted here. Briefly, adherent cells were detached by trypsinization, centrifuged and resuspended in complete media (these steps were omitted for the nonadherent CH27 cells), serially injected into a hollow glass capillary positioned between two optical fibers (Fig.~\ref{fig:OS}(a)), and  exposed to two 0.2\,W counterpropagating 1064\,nm laser beams to center each cell and allow it to rotate into an equilibrium orientation before stretching. Deformation was characterized by the edge-to-edge distance, along the laser axis, of a phase-contrast image of the cell, normalized to the distance measured during the 0.2\,W trapping period (Fig.~\ref{fig:OS}(a, inset)). In the current study this deformation was $\sim$1\% of the cell diameter.

In time-domain experiments, stretching power (0.9\,W per fiber, unless other specified, for 4\,s), and trapping power (0.2\,W per fiber for 2\,s)  were applied to  stretch the cell and allow recovery, respectively (Fig.~\ref{fig:OS}(b)). Simultaneously, cell images were recorded by phase contrast microscopy at 15--20\,frames\,s$^{-1}$. The photonic surface stress on a cell at the center of the beam, used to determine nominal creep compliance and complex modulus, was calculated via previously published models~\cite{guck2001optical} to equal approximately 0.3\,Pa per 1\,W laser power per fiber (see Supp.\ Info.) for the two optical stretching chambers used in this study.  

In frequency-domain experiments, cells were exposed to a sinusoidal laser profile for 8\,s with a mean power of 1\,W per fiber and a load amplitude of 0.5\,W per fiber (i.e., 1\,W peak-to-peak per fiber) unless otherwise specified, with a 1\,s trapping period at 1\,W per fiber before and after (Fig.~\ref{fig:OS}(c)). For oscillation frequencies of 10\,Hz and greater, a load amplitude of 1\,W per fiber was used. Images were recorded at 10--50 frames s$^{-1}$.  Amplitudes and phase angles were extracted from deformation signals by subtracting a moving average across one or more periods and fitting the expression $F\sin[\omega(t-t_0)-\phi]$ by nonlinear regression (Mathematica, Wolfram Research) where $F$ is the deformation amplitude, $\omega$ is the applied angular frequency, $t_0$ is the measured lag of the tool (collection, processing, and transmission time of laser data and image frames, see Supp.\ Info.) and $\phi$ is the phase angle. (Alternatively, fluidity and amplitude can be estimated by fitting the deformation to a quadratic function plus a sinusoid, with similar results; see Supp.\ Fig.~S\ref{fig:sine-fitting}.)  
 The signal-to-noise ratio was calculated by dividing the root-mean-square magnitude of the fitted sinusoid by the root-mean-square magnitude of the flattened deformation with the signal subtracted. During fixation experiments, 16\% of the chemically crosslinked cells exhibited signal-to-noise ratios $\mathrm{SNR}<1$ or unphysical values of fluidity $a<0$ or $a>1$; these cells were excluded from further analysis. 

Differences in structural-damping and lumped-component viscoelastic models are summarized in Supp.\ Info. In this work, optical stretching data was fitted to constitutive models of both types. The structural damping model in creep compliance took the form of $A(t/t_0)^a$ (with reference time $t_0=1$\,s) with different values of $A$ and $a$ used in stretching and recovery, yielding four parameters to fit. (Recovery was quantified as time-dependent contraction relative to the time and deformation at the end of the stretching period.) In this model, the phase lag is a frequency-independent $\phi=\pi a/2$. The lumped-component models contained four parameters, to offer equal complexity, and these were assigned to the two springs ($E_1$, $E_2$) and two dashpots ($\eta_1$, $\eta_2$) of a standard linear solid, which consists of a series-spring-dashpot pair ($E_1$, $\eta_1$) in series with a parallel-spring-dashpot pair ($E_2$, $\eta_2$). This model is most easily described by its deformation vs.\ load transfer function $D(s)=1/E_1+1/(s\eta_1)+1/(E_2+s\eta_2)$. The phase lag is $\phi(\omega)=\tan^{-1}\frac{\mathrm{Im}[D(i\omega)]}{\mathrm{Re}[D(i\omega)]}$. 

Optical stretcher operation provides cell diameter in the course of conducting an experiment, via video image capture and analysis. Cell diameter was found to be log-normally distributed with a geometric mean of 23\,$\upmu$m for the hMSCs and 18\,$\upmu$m for both the 3T3s and CH27s and a geometric standard deviation of approximately 1.1--1.2 for each cell type (Supp.\ Fig.~S\ref{fig:cell-sizes}), 
 where a geometric standard deviation of one would correspond to a single uniform cell diameter. The error (standard deviation) in repeated measurements of single cells was found to be 0.1\,$\upmu$m.

Error bars in all figures are standard error unless otherwise noted.

\subsection{Temperature characterization and control during optical stretching} 

In  optical stretching, laser beam absorption increases the temperature of the surrounding medium and trapped object (here, the cell); this temperature increase has previously been approximated as a constant value~\cite{ebert2006fluorescence,wetzel2011single,kiessling2013thermorheology}. Here we revise that estimate by developing a thermal model of time-dependent laser-induced heating (Supp.\ Fig.~S\ref{fig:supp-thermal-model}), 
 deriving a constitutive relationship that  follows an approximate $\ln t$ form: $T(t)=T_\infty+C_1\ln(1+C_2t)$, where $C_1$ and $C_2$  are constants representing the geometry and thermal characteristics of the system.
 
During external heating experiments, optical stretcher chamber temperature was controlled with two 10\,cm, $7\,\mathrm{W}\,\mathrm{cm}^{-2}$ strip resistance heaters that were clamped to the microscope stage. Cells were stored in a rotating syringe at room temperature~\cite{maloney2010} and were not exposed to elevated temperatures until they were injected into the capillary several seconds before stretching. A moving average was applied to external heating data to clarify trends; for this moving average only, we do not include the 13\% of cells in this data set with $\mathrm{SNR}<1$ or unphysical values of fluidity $a<0$ or $a>1$.

Temperature changes within microscale volumes were characterized by using the fluorescent dye Rhodamine B, the brightness of which is attenuated in a near-linear manner with increasing temperature~\cite{shah2009generalized}. From calibration experiments of dye intensity (with background subtracted) in an incubator microscope with adjustable temperature, we calculated an attenuation of 1.69\%\,$^\circ$C$^{-1}$ above room temperature $T_\infty=20\pm1^\circ$C. Dye brightness was insensitive to focal plane height, photobleaching was negligible when the dye was illuminated for several seconds only, and background fluorescent signal was easily measured by flushing  dye from the capillary. As a result, it was not necessary to use a reference dye such as Rhodamine 110.

\begin{figure*}
\centering
\includegraphics{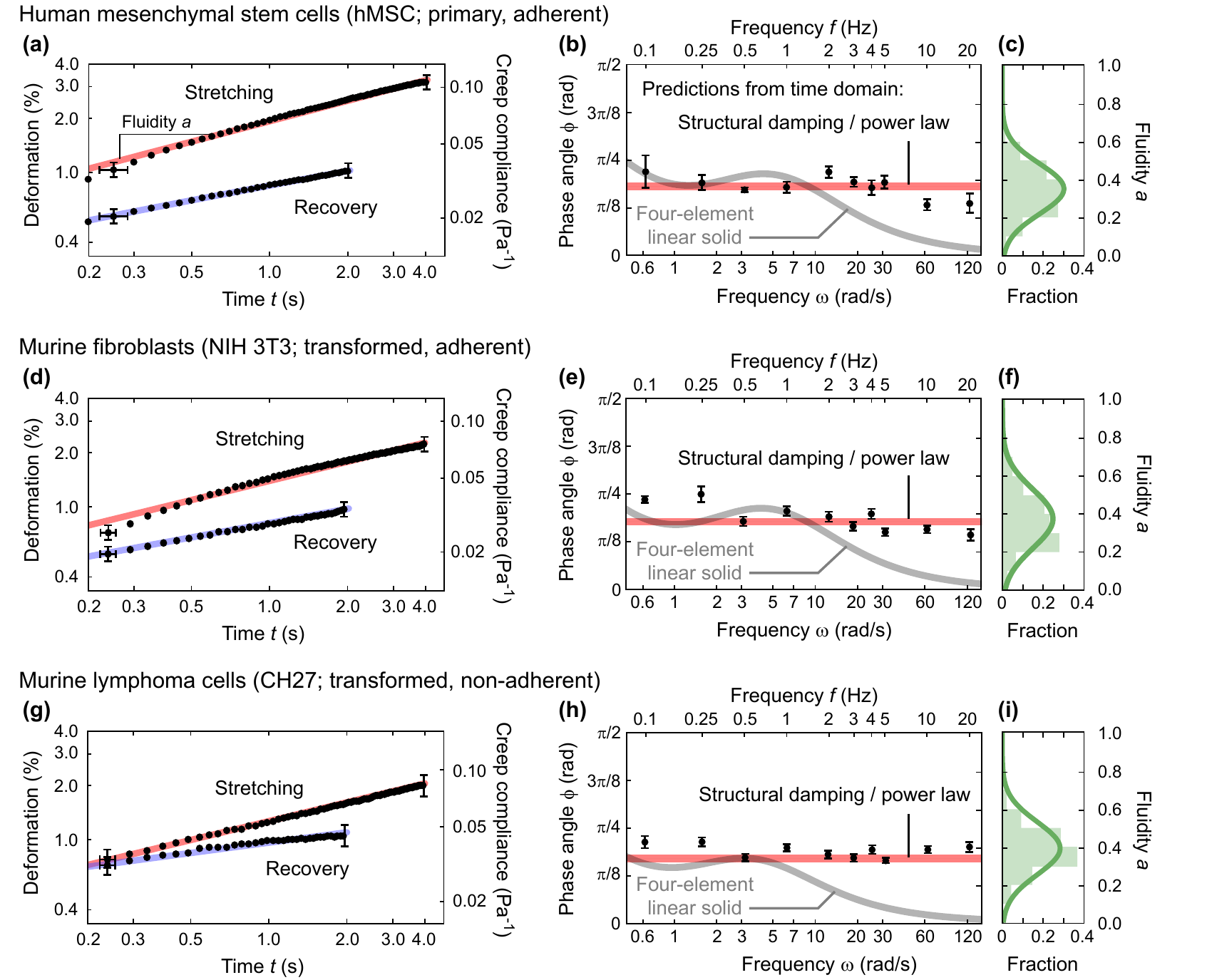}
\caption{Across three  eukaryotic cell types measured in the suspended state, structural damping (power-law) deformation behavior is consistent between the time and frequency domains at time scales  around 1\,s. (a,d,g)~In adherent and nonadherent primary and transformed cells, creep compliance  (geometric mean) in stretching and recovery is well-described by a power law that includes an exponent (here termed fluidity) and a multiplicative constant. (For clarity, error bars are shown for selected points only; hMSC creep compliance data from~\cite{maloney2010}. Average cell temperatures are estimated to be 40.1$^\circ$C and 28.5$^\circ$C during stretching and recovery, respectively.) A lumped-component model also containing four parameters can also be fit to the data (Supp.\ Fig.~S\ref{fig:spring-dashpot}). 
 (b,e,h)~Viscoelastic models fitted to time-domain data provide predictions of frequency-domain results that can be compared to measurements; here, the structural damping model predicts frequency-independent hysteresivity or fluidity, which is confirmed by measurements in the frequency domain. Equivalent-complexity four-element lumped-component models predict transitions in fluidity (corresponding to time constants of the spring-dashpot pairs) that are not observed in frequency-domain experiments and are thus apparently artifactual. (c,f,i)~Histograms show approximately Gaussian distribution of fluidity values. Fluidity estimates with standard error from all methods are tabulated in Table~S1.}
\label{fig:PLR}
\end{figure*}

\section{Results and discussion}

\subsection*{Structural damping / power-law rheology behavior is consistent across time and frequency domains, in contrast to spring-dashpot parameterization of equivalent complexity}

We conducted creep compliance (time-domain) analysis,  by applying unit-step laser-induced  photonic stresses on single whole cells from three eukaryotic cell populations: human mesenchymal stem cells (hMSCs), murine fibroblasts (3T3 FBs), and murine lymphoma cells (CH27s) (Fig.~\ref{fig:PLR}(a,d,g)), which constitute a selection of adherent and nonadherent and primary and transformed cell populations. The hMSCs are primary and adherent; the 3T3 fibroblasts are transformed and adherent; and the CH27 are transformed and nonadherent under typical \emph{in vitro} culture conditions. The full stretching and recovery periods generally resembled those shown in Fig.~\ref{fig:OS}(b); in Fig.~\ref{fig:PLR}(a,d,g) we show the recovery as a contraction relative to the time and deformation at the end of the stretching period.

The appearance of the  stretching and recovery data on a log-log scale of creep compliance $J(t)$ vs.\ time $t$ suggests a power-law fit ($J(t)\propto t^a$) with different exponents (i.e., different fluidity values) for the stretching and recovery periods, where $a\approx 0.4$ and 0.2, respectively. For comparison with a lumped-component model of equal complexity, we also fit a four-parameter standard linear solid to the creep compliance data for each cell type (Supp.\ Fig.~S\ref{fig:spring-dashpot}).
 Both models predict certain behavior in the frequency domain (e.g., the appearance or lack thereof of phase lag transitions at characteristic frequencies) that can then be compared to actual behavior in that domain.

To enable this comparison, we measured rheological parameters (stiffness and fluidity) in the frequency domain by applying sinusoidal photonic loads and recording the amplitude and phase lag of the sinusoidal portion of the resulting deformation (Fig.~\ref{fig:OS}(c)). Shown in Fig.~\ref{fig:PLR}(b,e,h) are the average fluidity measurements; histograms of these values are shown in Fig.~\ref{fig:PLR}(c,f,i). (We observed an approximate Gaussian distribution of fluidity, in agreement with previous studies of attached cells~\cite{desprat2005creep,balland2006power,hiratsuka2009power}. The fluidity of the primary hMSCs was independent of the number of population doublings since explantation and isolation; see Supp.\ Fig.~S\ref{fig:passaging}). 

Lumped-component models predicted time constants of approximately 0.1--3\,s (Supp.\ Fig.~S\ref{fig:spring-dashpot}), 
 (the viscosity-stiffness ratios of the spring-dashpot pairs, corresponding to characteristic frequencies of 0.06--1.6\,Hz) where one would expect fluidity to be altered strongly by changing frequency; however, these transitions were not observed (Fig.~\ref{fig:PLR}(b,e,h)). (We also did not see convincing evidence of a transition between multiple power laws~\cite{chowdhury2008cell,hoffman2009cell} over the frequency range of interest here, nor did we detect the onset of poroelastic, or decoupled solid--liquid, behavior that is expected to dominate at higher frequencies~\cite{fabry2001scaling,moeendarbary2013cytoplasm}.) Rather, fluidity was independent of frequency, within error, over two decades centered around 1\,Hz, and furthermore the magnitudes were in good agreement with those estimated by using the stretching portions of the creep compliance curves. The structural damping, or power-law, model was thus found to be viscoelastically consistent in the sense that predictions generated from the time domain were validated by measurements in the frequency domain. A lumped-component model of equal complexity consisting of springs and dashpots did not share this consistency, and therefore appeared to be not only a poorer fit, but also misleading at the time scale investigated here. (Our findings do not imply, however,  that cell deformation behavior could not be described accurately with more elements; power law rheology is equivalent to an infinite number of spring-dashpot assemblies with a power-law distribution of relaxation times~\cite{balland2006power}. Nevertheless, the principle of model parsimony argues against arbitrarily  large collections of lumped components~\cite{fabry2003time,lenormand2004linearity,puig-cytoskeletal}.)

We also acquired additional estimates of fluidity within the structural damping framework by fitting the complex modulus magnitude $|G^\star(\omega)|\propto \omega^a$. These estimates,  shown in Supp.\ Fig.~S\ref{fig:storage-modulus} 
 and  tabulated in Table S1, agreed with our other estimates. Taken as a whole, the results reinforce our earlier conclusion~\cite{maloney2010} that the stress fibers often observed in the attached state are not a necessity for manifestation of structural-damping or power-law rheological behavior.  
In summary, we reject a lumped-component parameterization of several springs and dashpots not only due to poor performance by fitting metrics~\cite{maloney2010} (see also~\cite{roca2006rheology,hemmer2009role,zhou-power}), but also due to inconsistency between predictions in the time and frequency domains. That fundamental inconsistency, under the reasoning that extracted time constants simply reflect the experiment duration, has itself been predicted~\cite{fabry2003time} but, to our knowledge, not previously demonstrated.

\subsection*{Fluidity modulated by applied load}

\begin{figure}
\centering
\includegraphics{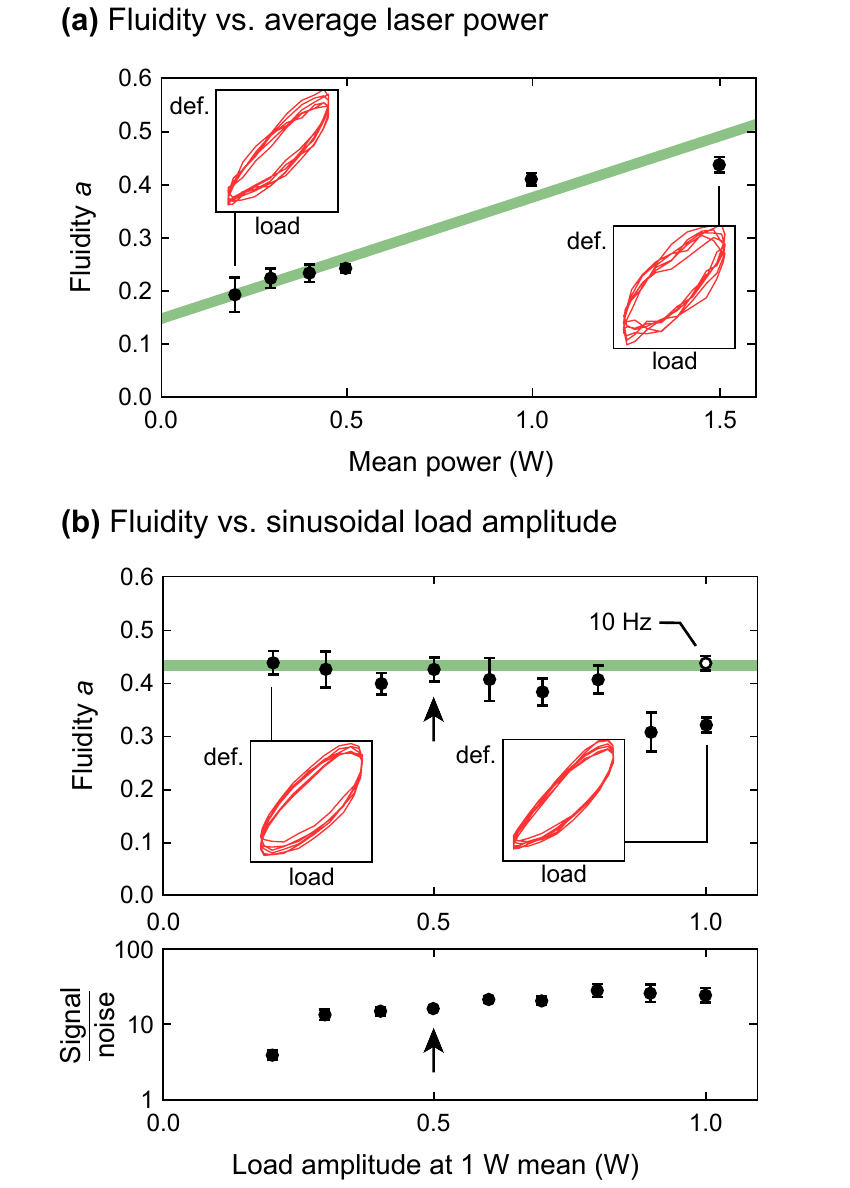}
\caption{Effects of thermomechanical loading on suspended-cell fluidity. (a)~In frequency-domain optical stretcher experiments, fluidity increases with mean incident laser power ($n=40$--324 cells per power setting). (b)~At a mean power of 1\,W per fiber, estimated mean cell fluidity is minimally dependent on load at sufficiently low load amplitudes ($n=10$--165 cells per power setting). Below, increase in signal-to-noise ratio above 10 for load amplitude of  0.3\,W per fiber and above suggests a window for obtaining relatively low-noise data without drastically changing the parameter of interest through loading. (Arrows show 0.5\,W amplitude setting used for chemical fixation and ATP depletion investigations.)  Inset, symmetric and elliptical Lissajous figures show that cells do not exhibit noticeably nonlinear effects such as strain stiffening or softening at the settings used.}
\label{fig:linearity}
\end{figure}

To determine the sensitivity of fluidity values  against changes in applied photonic load, we stretched  cells in the frequency domain (1\,Hz) with multiple mean and  amplitude laser power values. We observed fluidity values to increase with increasing mean laser power (Fig.~\ref{fig:linearity}(a)).  Additionally, a sufficiently large load amplitude caused fluidity to deviate detectably from a constant small-load-amplitude value (Fig.~\ref{fig:linearity}(b)); a possible origin for this decrease is discussed in Supp.\ Info. Smaller load amplitudes, in contrast,  were correlated with a relatively small signal-to-noise ratio, such that it was difficult to discern signals arising from load amplitudes of less than 0.2\,W per fiber, at least for the 1\,Hz, 8\,s stretching settings employed here. We thus selected a sinusoidal amplitude of 0.5\,W per fiber (overlaid on a constant power of 1\,W per fiber) to evaluate the influence of chemicals on cell fluidity, as discussed in the next section. This condition provided a suitable range of load amplitudes that produced sufficiently high signal-to-noise ratios from deformation signals while minimally altering the extracted magnitudes of fluidity. Note that, for all load amplitudes employed, Lissajous figures of load vs.\ deformation remained generally elliptical and symmetric. In our observations, all whole-cell deformations (with magnitudes of up to 2\% and deformation rates of up to 20\%\,s$^{-1}$) remained in the linear viscoelastic regime (for comparison, see reports of mechanical nonlineararity for eukaryotic or red blood cells under other loading conditions~\cite{fernandez2006master,fernandez2008single,puig2007viscoelasticity}).

\begin{figure}
\centering
\includegraphics{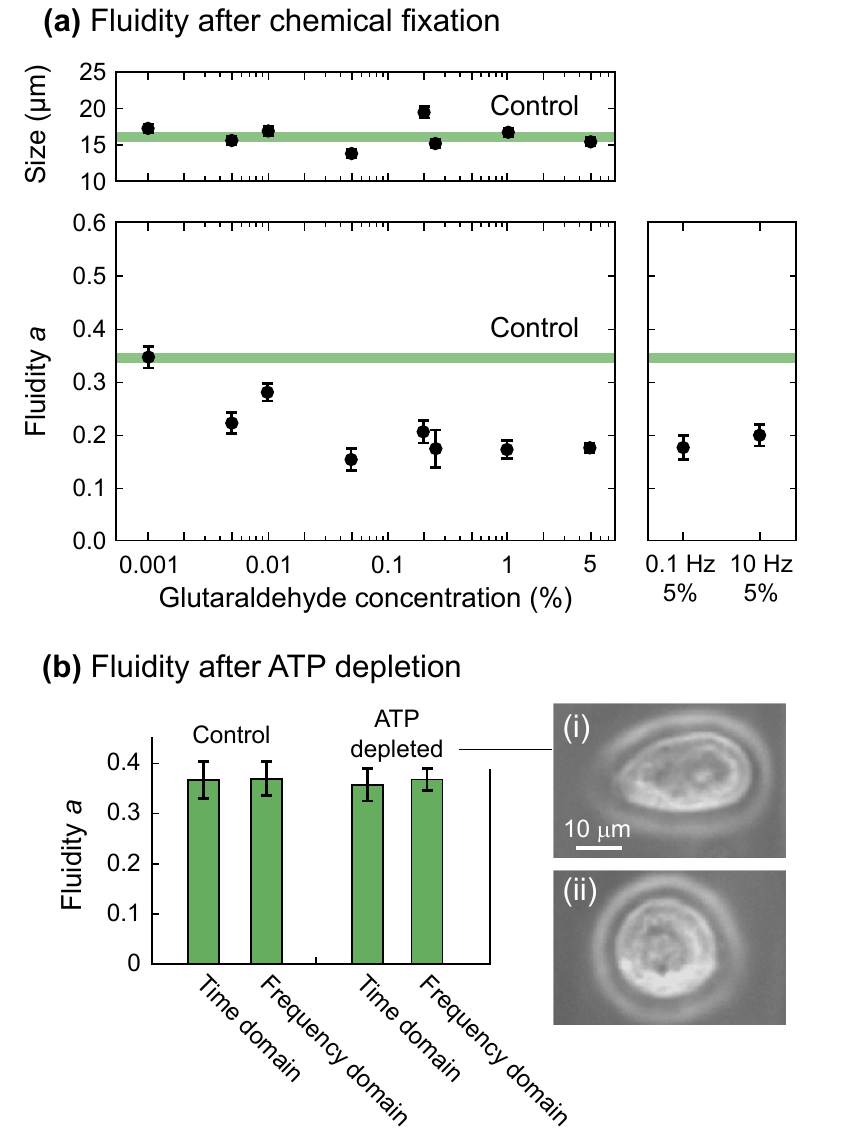}
\caption{Suspended-cell fluidity is reduced by chemical fixation, but not detectably altered by ATP depletion. (a)~Glutaraldehyde fixation reduces fluidity in a dose-dependent but frequency-independent manner, in agreement with previous reports that chemical crosslinking hinders network rearrangement that enables fluidity in the cell, but cell diameter is unaffected ($n=11$--123 CH27 cells per concentration, $n=29$ cells for control). (b)~Fluidity is consistent between time- and frequency-domain measurements when ATP is depleted, similar to controls (untreated cells) ($n=102$--249 3T3 fibroblasts per condition and domain). Inset, photographs of cells having undergone ATP depletion (i) before and (ii) after trypsinization show that contractile machinery is hindered after depletion and that the cell is unable to remodel from an attached-state to a suspended-state morphology.)}
\label{fig:chemicals}
\end{figure}

\subsection*{Impact on fluidity of chemical fixation and ATP depletion} 

We investigated chemical perturbation of fluidity with two approaches: glutaraldehyde fixation, which covalently crosslinks the cytoskeleton and kills the cell in the process; and ATP depletion, which generally slows metabolic processes and blocks  actomyosin contraction to possibly alter whole-cell deformation. We extracted fluidity values from  frequency-domain (phase lag) measurements of CH27 cells exposed to various glutaraldehyde concentrations for 10 min at 37$^\circ$C. Figure~\ref{fig:chemicals}(a) illustrates a dose-dependent reduction in fluidity $a$ from 0.35 to 0.18 with crosslinking extent. (Note that increased incubation time did not detectably reduce fluidity further, $a=0.17\pm 0.02$ for 30\,min vs.\ $a=0.18\pm 0.01$ for 10\,min.) 

Experiments at 0.1\,Hz and 10\,Hz showed that the reduction of fluidity occurred evenly across this frequency range. Insensitivity of cell size to glutaraldehyde treatment excluded volumetric changes as an origin for fluidity alteration. The gradual transition over at least three decades of fixative concentration represents a decrease in the rate of cytoskeletal rearrangement processes, as transient protein-protein interactions are replaced by covalent bonds~\cite{chowdhury2008cell}. This modulation of fluidity via crosslinking constitutes a positive control for tracking changes in this parameter, to be compared to similar reductions in fluidity by similar treatment of attached cells~\cite{chowdhury2008cell}.

It has been suggested that adenosine triphosphate (ATP) hydrolysis enables a nonzero fluidity value in cells~\cite{bursac2005cytoskeletal,laudadio2005rat,fredberg2006cytoskeleton,trepat2007universal}. Previous investigations of ATP depletion effects on (power-law) rheology have been reported for cells in the attached state (e.g.,~\cite{bursac2005cytoskeletal,hoffman2006consensus,chowdhury2008cell}); we here test this hypothesis with cells in the suspended state in which ATP synthesis is inhibited and ATP stores are depleted by chemical means. Interestingly, ATP depletion in the present experiments, confirmed by chemical assay and cell morphology (Fig.~\ref{fig:chemicals}(b(i,ii))), did not alter whole-cell mean fluidity $a$, within error, in 3T3 fibroblasts. (See Fig.~\ref{fig:chemicals}(b), where fluidity values are obtained from, and are in agreement between, both time- and frequency-domain measurements.) Equivalent results were obtained from CH27 cells, $a=0.39\pm 0.02$ after ATP depletion vs.\ $a=0.39\pm 0.01$ for control, via frequency-domain testing of 25 cells per condition.

These findings have important implications when investigating the origin of nonzero fluidity that is also conserved over multiple frequency decades (Figs.~\ref{fig:PLR},~\ref{fig:chemicals}(a)). The fluidity $a$ of soft glassy materials has been proposed to represent (through a ``noise temperature'' equivalent to $a+1$)  an effective mean-field energy ($\gg$$kT$) that describes jostling from the mechanical yielding of neighboring regions in the material~\cite{sollich1998rheological}. It has been in turn noted that ATP is an effective energy carrier that the cell employs for actomyosin contraction, leading to the proposal that ATP hydrolysis provides the agitation needed to enable power-law rheology in cells~\cite{fredberg2006stress,nguyen2008strange,trepat2008universality,kollmannsberger2011linear}. However, others have reported from experiments that in adherent, contractile cells, fluidity is minimally altered even when ATP is depleted~\cite{bursac2005cytoskeletal,hoffman2006consensus,van2006role,trepat2007universal,chowdhury2008cell}.  Now shown in suspended cells as well, we find that the absence of ATP leaves the cytoskeleton in a rigor state that nevertheless remains susceptible to other sources of structural jostling or agitation and---like inanimate soft glassy materials---continues to exhibit  power-law rheology with nonzero fluidity. 

\subsection*{Fluidity increases with temperature}

\begin{figure}
\centering
\includegraphics{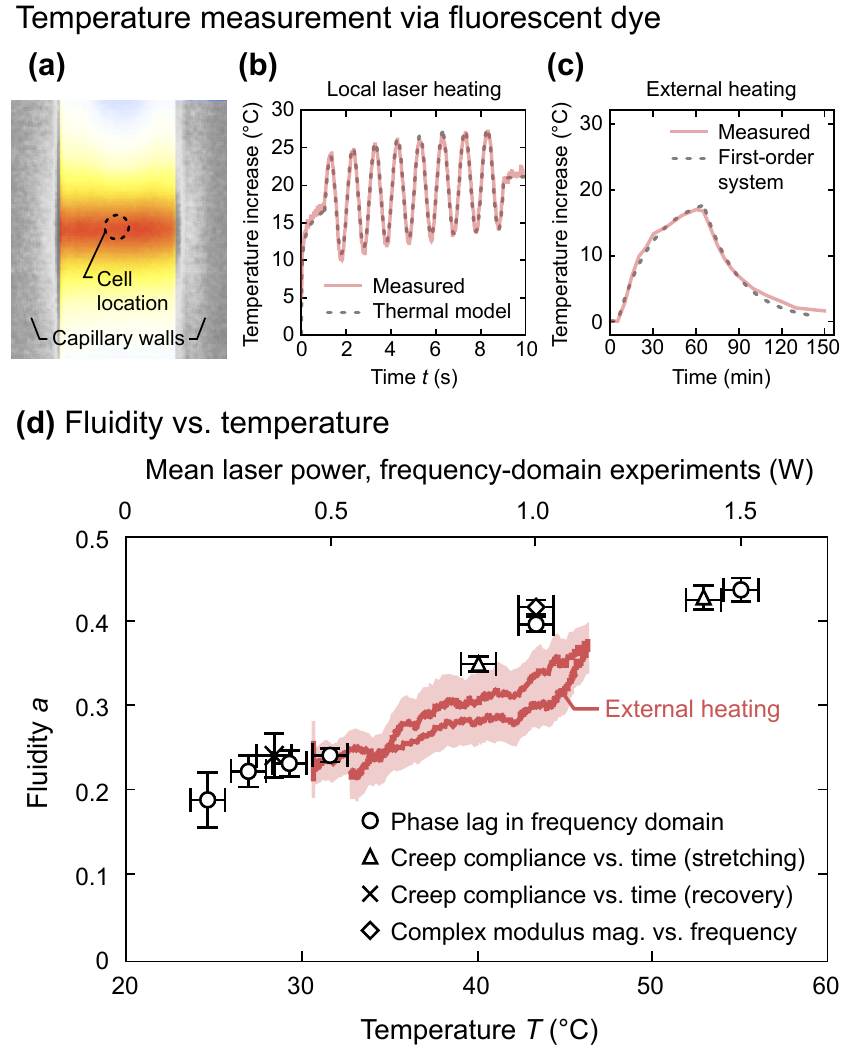}
\caption{Whole-cell fluidity increases with temperature, according to optical stretching measurements in the time and frequency domains, in which temperature was controlled via external heating and/or laser absorption. (a)~A temperature-sensitive fluorescent dye allows local laser-induced heating to be quantified. (b)~A temperature increase with a $\ln(t)$ form over time $t$ in response to a laser power unit step is derived (Supp.\ Fig.~S\ref{fig:supp-thermal-model}) 
 and used to predict temperature response from arbitrary laser power profiles. (c)~Resistance heaters clamped to the microscope stage enable external heating (with the thermal characteristics of a first-order system), decoupling cell temperature from photonic load. (d)~Enabled by fluorescent-dye temperature quantification and fluidity obtained from different rheological techniques ($J(t)$ vs.~$t$, $G(\omega)$ vs.~$\omega$, and phase lag $\phi$) at different laser powers, along with external heating, fluidity is found to increase with temperature at an approximate rate of 0.01\,$^\circ\mathrm{C}^{-1}$. External heating values show 100-cell moving average of fluidity from 860 individual cells with 95\% confidence interval during the heating and cooling cycle shown in (c).}
\label{fig:fluidity}
\end{figure}

Finally, we extend the relatively small number of studies addressing temperature-dependent cell rheology~\cite{bursac2005cytoskeletal,picard2009impact,sunyer2009temperature,kiessling2013thermorheology}. Temperature adjustment provides a powerful tool to characterize viscoelastic materials and to evaluate models of such materials, and photonic tools such as the optical stretcher provide a means to alter local temperature rapidly~\cite{peterman2003laser,ebert2006fluorescence,wetzel2011single}. Because laser power is coupled to both increased applied stress and to increased temperature of cells (the second effect being an unintended and generally undesirable consequence of irradiation during optical stretching~\cite{wetzel2011single}), we decoupled these cues by externally heated the optical stretcher chamber with resistance heaters attached to the microscope stage while the cells were injected into the chamber from an off-stage syringe. 

We began by characterizing the temperature increase from laser heating alone. The temperature-sensitive fluorescent dye Rhodamine B was used to measure the optical stretcher chamber temperature during these experiments; heat map coloration of the resulting fluorescence intensity allows visualization of the temperature field (Fig.~\ref{fig:fluidity}(a))~\cite{ebert2006fluorescence}.  In Supp.\ Info., we develop a thermal model of the time response of temperature changes in response to laser heating; a comparison of predicted and measured temperature changes is shown in Fig.~\ref{fig:fluidity}(b). When set to dissipate approximately 20\,W\,in$^{-2}$, the stage-mounted external heaters raised the capillary and medium temperature by several tens of degrees in the manner of a first-order thermal system with a time constant of approximately half an hour, verified by dye measurements (Fig.~\ref{fig:fluidity}(c)).

Following this characterization of laser-induced and external heating, we stretched CH27 cells for one hour, then heated and (passively) cooled the stage with the temperature profile shown in Fig.~\ref{fig:fluidity}(c)  by turning the heaters on and off at 60 and 120 minutes, respectively. We stretched cells by using a 1\,Hz sinusoidal laser power amplitude of 0.5\,W per fiber  overlaid on a constant power of 0.5\,W per fiber. As a result, cells were stretched at average temperatures ranging from 31 to 47$^\circ$C, providing individual fluidity values cross-referenced to cell temperature (Supp.\ Fig.~S\ref{fig:supp-fluidity-vs-time}). 
 As shown in Fig.~\ref{fig:fluidity}(d), a 100-cell moving average over  \textgreater 800  cells displays a trend of increasing fluidity with increasing temperature, and vice versa, during this cycle of external heating. 

Also shown in Fig.~\ref{fig:fluidity}(d) are summarized fluidity measurements acquired from CH27 cells in the time and frequency domains, including the data shown in Fig.~\ref{fig:PLR}(g,h,i). This collection of results, now shown as a function of cell temperature, enables two important conclusions. First, the increase in fluidity with laser power is linked with laser-induced heating, as the trend observed with external heating is in good agreement with that observed with laser-induced heating. Second, the different fluidity values obtained during creep compliance stretching and recovery can now be interpreted as a single, temperature-dependent parameter. (These conclusions also hold  for the two other cell types studied here; see Supp.\ Fig.~S\ref{fig:supp-fluidity}.)

We conclude that for measurements at different temperatures, it is appropriate to use the more general constitutive equation $G^\star(\omega,T)=g_0(T)(i\omega/\omega_0)^{a(T)}$ for a fixed temperature, or a convolution approach, for example, when the temperature changes during an experiment, as is the case with optical stretching. (The relationship between whole-cell stiffness $g_0(T)$ and temperature $T$ is shown in Supp.\ Fig.~S\ref{fig:supp-stiffness-vs-temp}, 
 where  creep compliance values were converted to stiffness $g_0$ in $G^\star(\omega)=g_0(i\omega/\omega_0)^a$ by using  $g_0=[J(1\,\mathrm{s})(2\pi)^a\Gamma(1+a)]^{-1}$ where $\Gamma$ is the gamma function~\cite{balland2006power}. Here also, results from external heating and laser-induced heating are in good agreement.) This interpretation supersedes the offset-power-law constitutive relation that we considered previously~\cite{maloney2010}.

Such findings clarify a relationship that has rarely been examined---and even then has been complicated by the influence of contraction-induced prestress when fluidity is measured in attached cells. For example, Bursac et al.\ reported a monotonic increase in fluidity (or equivalently, power-law exponent) with temperature~\cite{bursac2005cytoskeletal}, for cells in the attached state. In contrast, Sunyer et al.\ measured a decrease in fluidity with increasing temperature~\cite{sunyer2009temperature}, which (with the use of pharmacological inhibition) was attributed to substratum-prompted contraction, an effect absent in the suspended state. Kie\ss ling et al.\ recently studied the thermorheology of single suspended cells at deformation times near 1\,s, via an optical stretcher modified with a second set of fibers, under the framework of simple time-temperature superposition (TTS)~\cite{kiessling2013thermorheology}. Note that simple TTS is not strictly compatible with a power-law material (Fig.~\ref{fig:PLR}) that exhibits temperature-dependent fluidity (Fig.~\ref{fig:fluidity}(d)), because constant time and deformation scaling factors cannot reproduce a creep compliance relationship of $J\propto t^{a(T)}$. As we describe in Supp.\ Info., however, if fluidity is assumed to be  independent of temperature over small ranges, one can derive a time scaling parameter in agreement with Kie\ss ling et al. It may ultimately be a matter of convenience whether one employs the approximation of simple TTS or not; in the current work, we prefer to express the rheological behavior in terms of temperature-dependent fluidity and stiffness as a description of a linear viscoelastic material.

\section{Conclusion}

We expect that physical sorting approaches will continue to emerge that will test suspended cells at time scales of approximately 1\,s or smaller, and that studies of single cell mechanics will provide robust parameterization while enabling theories of soft matter and complex fluids to be tested. Within this scope, our goals are to characterize the mechanical behavior of fully suspended cells, to extend familiarity with animate soft condensed matter and the origins of deformation, and to enable interpretation of cell deformation in cell-sorting devices to separate cell populations by leveraging deformation mechanisms relevant at the time scale of interest. The term ``fluidity'' generalizes a parameter that can be estimated from phase lag or power-law exponent (with the most robust results obtained here by using phase lag), without committing to any specific model of cell deformation and without the need to calibrate stiffness and compliance to absolute values---a special advantage with optical stretching, where optical coupling means that deformation amplitude varies with refractive index and individual chamber geometry.

We thus devoted part of the current study to examine the advantages (and the potential pitfalls) of applying one viscoelastic framework or another. While there is no guarantee that fitted models will extrapolate accurately outside the range to which they are fit, and while we therefore emphasize that our parameterization is applicable specifically to suspended whole cells at the time scale studied here, it is certainly reasonable to expect fitted models to be internally consistent between time and frequency domains. However, we found that time constants predicted from lumped-component fits in the time domain were not observed in the frequency domain. Spring-dashpot models of equivalent complexity therefore appear to be viscoelastically inconsistent between time and frequency domains, producing predictions of system time constants that are apparently artifactual. 

Following these conclusions, we examined fluidity as a parameter with which to characterize cells and cell populations and the effects of chemical and environmental perturbations. We tested the hypothesis that ATP hydrolysis is specifically the source of athermal agitation that enables structural damping or power-law rheology in cells. However, we found that ATP depletion (visibly confirmed to inhibit remodeling in recently suspended cells), did not transform suspended cells into elastic solids or even alter their fluidity within error. Therefore, ATP appears to not be crucial in enabling power-law rheology during linear viscoelastic deformation of cells, though ATP hydrolysis is plausibly linked to cytoskeletal network remodeling after large fluidizing deformations. It is emphasized that this conclusion needs to be expressed precisely, as ATP plays a part in cytoskeletal resolidification after disruption, and it is thus too vague to say that ATP hydrolysis is or is not responsible for power-law rheology. The present study's findings are specifically that ATP hydrolysis played no detectable part in the average fluidity or hysteresis (as quantified by phase lag) during oscillatory measurements in the linear regime.

The temperature dependence of fluidity in the suspended cell has now been established. For sufficiently small load amplitudes, fluidity was essentially constant; however, we found that its value increased  with the mean photonic power used to deform the cell. We decoupled laser-induced stress and temperature to show that whole-cell fluidity depended on temperature, a  finding relevant for biophysical understanding of the cell and its position in the framework of animate and inanimate soft condensed matter. A collection of fluidity estimates from the time and frequency domains contributes to a quantitative understanding of temperature dependence and further explains rheological differences between stretching and recovery portions of creep compliance data. Optical stretching changes the temperature of the cell, specifically between stretching and  and recovery  segments of creep compliance experiments; it is now possible to explain the difference in fluidity estimates during these segments as manifestations of a general temperature dependence. 

In summary, the fluidity $a$ of cells can be measured rigorously and compared as a function of cell type, chemical state and physical environment---without potential artifacts of cell-substratum and cell-tool contact. Measurement of this nondimensional rheological parameter suggests that the structural damping model best predicts cell mechanical response. Such parameterization supports rapid and accurate assays of cell mechanics in contexts relevant to sorting, delivery, and study of suspended cells.

\begin{acknowledgments}
This work was supported by the Singapore-MIT Alliance for Research and
Technology (SMART) Centre (BioSyM IRG), NSF CAREER CBET-0644846  (KJVV), NSF REU DBI-1005055 (EL \& AFL), and the NIH/NIBIB Molecular, Cellular, Tissue and Biomechanics Training Grant EB006348  (JMM). We gratefully acknowledge guidance from J.~Guck et al.\ (Cambridge University and Technical University of Dresden) on optical stretcher construction and image analysis. We appreciate the donation of CH27 lymphoma cells by S.~H.~Um and D.~J.~Irvine (MIT).
\end{acknowledgments}

\section{Supporting Citations}
References~\cite{lincoln2006phdthesis,poularikas2009transforms,hale1973optical} appear in the Supporting Material.

\bibliography{Fluidity}
\clearpage

\setcounter{figure}{0}

\section*{Supplemental Information: Viscoelastic framework and calculation of laser stress}

In viscoelastic materials,  stress $\sigma(t)$ and  strain $\varepsilon(t)$ are generally out of phase, with the complex modulus $G^\star(\omega)=G^\prime(\omega)+iG^{\prime\prime}(\omega)$ in the linear regime calculated as  $\sigma(t)/\varepsilon(t)$ where $G^\prime$ is the storage modulus and $G^{\prime\prime}$ is the loss modulus. The complex modulus magnitude or stiffness $|G^\star(\omega)|$ is calculated as $\sigma_0/\varepsilon_0$, where $\sigma_0$ is the amplitude of the photonic sinusoidal stress and $\varepsilon_0$ is the amplitude of cell sinusoidal deformation. The creep compliance $J(t)$ is $\varepsilon(t)/\sigma_0$ following a unit stress step $\sigma_0$. Fluidity $a$ is calculated as $a=(2/\pi)\tan^{-1}(G^{\prime\prime}/G^\prime)=2\phi/\pi$, where $\phi$ is the phase lag of sinusoidal cell deformation in units of radians. 

The viscoelastic response of whole, single cells around the time scale of 1\,s is commonly  fit to a structural damping (power-law) or lumped-component (spring-dashpot) model, with exemplary load-deformation functions given in the main text. In the structural damping model, this fluidity is independent of frequency and is numerically equal to the exponents appearing in the creep compliance $J(t)=j_0(t/t_0)^a$ and the complex modulus  $G^\star(\omega)=g_0(i\omega/\omega_0)^a$, where $t_0=1\,\mathrm{s}$ and $\omega_0=1\,\mathrm{rad}\,\mathrm{s}^{-1}$. Lumped-component models are assemblies of connected springs and dashpots, which individually exemplify purely Hookean-solid elastic and Newtonian-liquid viscous response, respectively. These models predict frequency-dependent fluidity; specifically, spring-dashpot pairs exhibit time constants that correspond to fluidity transitions; for example, a series-spring-dashpot pair transitions from an elastic stiffness $E$ to a fluid viscosity $\eta$ ($a=0\rightarrow 1$) around a central frequency $E/\eta$, while a parallel-spring-dashpot pair undergoes the opposite transition. 

Note that in a previous study of hMSCs in the time-domain~\cite{maloney2010}, we employed an offset power law (with exponent $a=0.25$) to represent cell behavior in both stretching and recovery; that offset power law scored better than a pure power law (with $a=0.37$) for hMSCs according to fitting metrics applied to  data available at that time (supporting material in~\cite{maloney2010}). However, frequency-domain measurements of cells presented here  strongly support a pure power-law model, which we now employ. We thus consider two different power-law exponents for creep compliance stretching and recovery, and  look explicitly at the role of temperature in causing these fluidity values to be different.

In optical stretching,  optical fibers emit unfocused laser beams that diverge and widen   through index-matching gel,  the glass wall of a square capillary, and finally liquid saline or cell growth medium before reaching the cell. The photonic stress $\sigma_0$ on the cell surface at the laser axis during optical stretching is calculated as~\cite{guck2001thesis,lincoln2006phdthesis}
\begin{subequations}
\begin{equation}
\sigma_0=\frac{2P(n_\mathrm{cell}-n_\mathrm{med})}{c\pi w_\mathrm{beam}^2n_\mathrm{med}},
\end{equation}
\begin{equation}
w_\mathrm{beam}=w_\mathrm{core}\sqrt{1+\frac{B^2}{z_0^2}},
\end{equation}
\begin{equation}
B=n_\mathrm{gel}\left(\frac{z_\mathrm{med}}{n_\mathrm{med}}+\frac{z_\mathrm{glass}}{n_\mathrm{glass}}\right)+z_\mathrm{gel},
\end{equation}
\begin{equation}
z_0=\frac{w_\mathrm{core}^2\pi n_\mathrm{gel}}{\lambda},
\end{equation}
\end{subequations}
where $P$ is the laser power per fiber; $n_\mathrm{cell}=1.372$, $n_\mathrm{med}=1.335$, $n_\mathrm{gel}=1.449$, and $n_\mathrm{glass}=1.474$ are the cell, medium, gel, and glass refractive indices, respectively; $z_\mathrm{med}=40\,\upmu$m, $z_\mathrm{glass}=20\,\upmu$m, $z_\mathrm{gel}=90$ or $106\,\upmu$m (depending on the chamber used) are the nominal medium, glass, and gel distances, respectively; $w_\mathrm{core}$ is a radius of 3.1\,$\upmu$m for the fibers used here; $c$ is the speed of light, and $\lambda=1064$\,nm is the laser wavelength.

\clearpage
\section*{Supplemental Information: Calculation of tool lag and material phase lag}

When extracting a phase lag from the signals acquired from edge detection of phase contrast images, we found it essential to account for systematic delays caused by communication between the  operating software (National Instruments LabView) and the data acquisition (DAQ) card (National Instruments USB-6229), the DAQ card and the laser (IPG Photonics YLR-3X2-1064); and the camera (Allied Vision Technology Marlin 146B) and the operating computer's Firewire communications port. These delays added up to a detectable and sometimes considerable fraction of the phase lag caused by the viscoelastic nature of the cell. The characteristic time delay of the system was estimated by fitting a sinusoid to the measured brightness caused by a sinusoidal laser profile with the IR filter removed. The delay is $t_0=24.9\,\mathrm{ms} + ROI(0.086\,\mathrm{ms})+1/(2r)$, where $ROI$ is the region of interest height in pixels (typically 160--200) after 2$\times$2 binning of the image collected from the camera and $r$ is the frame rate, with an uncertainty (standard deviation) of $e = 0.3$\,ms based on multiple tests. This uncertainty introduces an error in the fluidity of $4fe$ that is generally insignificant at frequency $f<10$\,Hz, growing to 0.024 at 20\,Hz. This error is included in Fig.~\ref{fig:PLR}(b,e,h).

\clearpage
\section*{Supplemental Information: Estimate of fluidity from fitting}

\begin{figure}[h]
\centering
\includegraphics{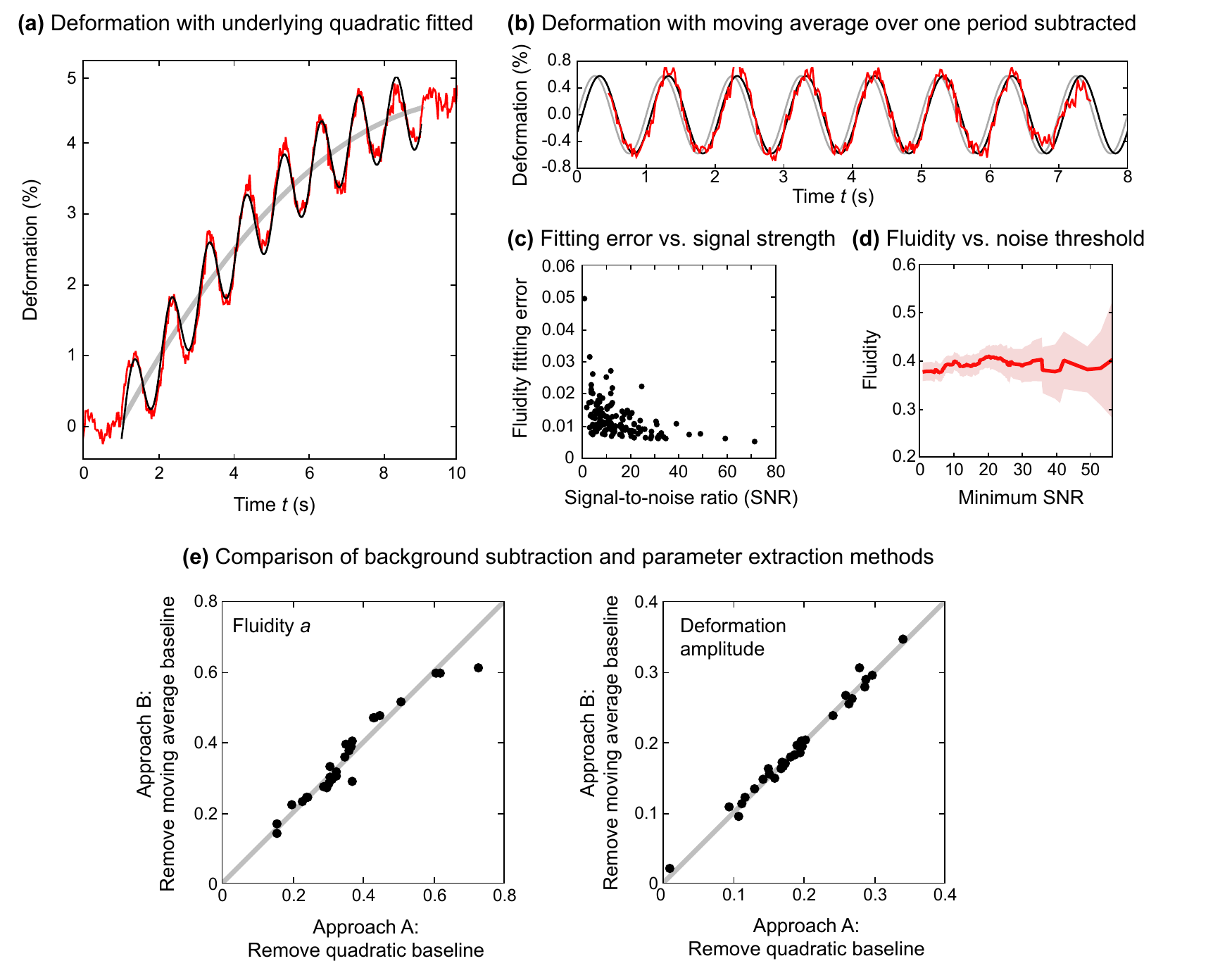}
\caption{(Supplemental) Two techniques were explored for estimating sinusoidal phase lag superposed with creep in a linear viscoelastic system. (a)~In the first approach, background deformation is fit to a quadratic function. (Estimates: fluidity $a= 0.27$, 0.56\% deformation amplitude.) (b)~In the second approach, a moving average over one period is subtracted to leave the sinusoidal component. (Estimates: fluidity $a= 0.28$, 0.58\% deformation amplitude. The signal-to-noise ratio for this signal is 16.2.) (c,d)~When obtaining fluidity by nonlinear fitting,  error  at the individual cell level is inversely related to signal-to-noise ratio (SNR); average fluidity across a cell population is  insensitive to minimum SNR. (e)~From a collection of 30 cells, comparison of estimates from both methods shows good agreement; the second approach was generally used for parameter estimation in this study, with the first approach used as a check for consistency between the two approaches.}\label{fig:sine-fitting}
\end{figure}

\clearpage
\section*{Supplemental Information: Cell size}

\begin{figure}[h]
\centering
\includegraphics{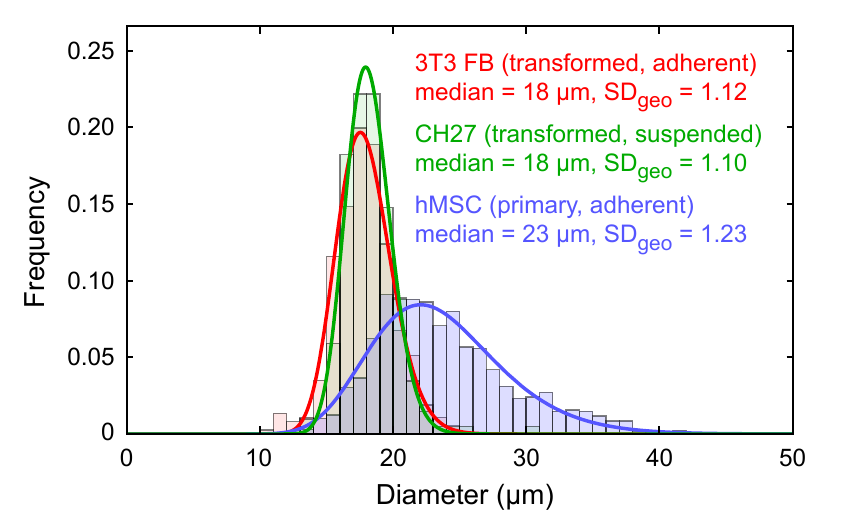}
\caption{(Supplemental) Distribution of cell diameter of three types of fully suspended cells, acquired during optical stretching.}\label{fig:cell-sizes}
\end{figure}

\begin{table}
\centering
\caption{(Supplemental) Estimates of fluidity (a measure of hysteresivity normalized to solid and liquid extremes, independent of frequency under the structural damping parameterization) and fluidity uncertainty for three eukaryotic cell types in the suspended state, as measured by optical stretching in the time and frequency domains. Fluidity differences at different optical stretcher laser powers are attributed to this parameter's temperature dependence; see Fig.~\ref{fig:fluidity}(d).}
\begin{footnotesize}
\begin{tabular}{lccc}
\hline
& ~Human mesenchymal~ & ~Murine fibroblasts~ & ~Murine lymphoma~ \\
Parameter & stem cells (hMSCs) &  (NIH 3T3s) &  cells (CH27s) \\ \hline
\multicolumn{4}{l}{From creep compliance  (geometric mean) vs.\ time (Fig.~\ref{fig:PLR}(a,d,g)):} 
 \\ \rule{0pt}{3ex}\textbf{Fluidity (stretching, 0.9\,W/fiber)} & \textbf{0.38} & \textbf{0.36} & \textbf{0.34} \\
Standard error$^*$ & \textless 0.01~~\, & \textless 0.01~~\, & 0.02 \\
\rule{0pt}{3ex}\textbf{Fluidity (recovery, 0.2\,W/fiber)} & \textbf{0.25} & \textbf{0.24} & \textbf{0.24} \\
Standard error$^*$ & \textless 0.01~~\, & 0.01 & 0.03 \\
\rule{0pt}{3ex}Data set size & 1248 & 288 & 208 \\
\hline\multicolumn{4}{l}{From sinusoidal phase lag (Fig.~\ref{fig:PLR}(c,f,i)):} 
 \\ \rule{0pt}{3ex}\textbf{Average fluidity, 1\,W/fiber} & \textbf{0.35} & \textbf{0.38} & \textbf{0.38} \\
Standard deviation$^{**}$ & 0.13 & 0.17 & 0.14 \\
Standard error & \textless 0.01~~\, & \textless 0.01~~\, & \textless 0.01~~\, \\
\rule{0pt}{3ex}Data set size & 373 & 433 & 1046 \\ 
\hline\multicolumn{4}{l}{From complex modulus (geometric mean) vs.\ frequency (Fig.~S\ref{fig:storage-modulus}):} 
 \\ \rule{0pt}{3ex}\textbf{Fluidity, 1\,W/fiber} & \textbf{0.50} & \textbf{0.49}  & \textbf{0.41} \\
Standard error$^{***}$ & 0.05 & 0.03 & 0.01 \\
\rule{0pt}{3ex}Data set size & 183 & 99  & 633 \\ \hline
\multicolumn{4}{l}{$^*$Standard error of best-fit line obtained by bootstrapping to convergence~\cite{maloney2010}.}\\
\multicolumn{4}{l}{$^{**}$Standard deviation of individual fluidity values.}\\
\multicolumn{4}{l}{$^{***}$Standard error of best-fit line obtained by linear regression.}\\

\label{table:data}
\end{tabular}
\end{footnotesize}
\end{table}

\clearpage
\section*{Supplemental Information: Temperature increase during optical stretching}

In optical stretching, photonic stress is used to deform cells with the advantage of completely avoiding physical contact. However, the laser beams are absorbed to some extent by the cell and the surrounding medium (generally  saline or cell growth medium), causing heat generation. 
As a result, the cell is brought to an elevated temperature in the process of being deformed. This temperature increase has been previously characterized by temperature-sensitive fluorescent dye and approximated as an instantaneous unit step to a constant final value~\cite{ebert2006fluorescence,wetzel2011single,kiessling2013thermorheology}. Here, we extend this previous work by deriving and confirming a model of laser-induced temperature increase over time.

\begin{figure}
\centering
\includegraphics{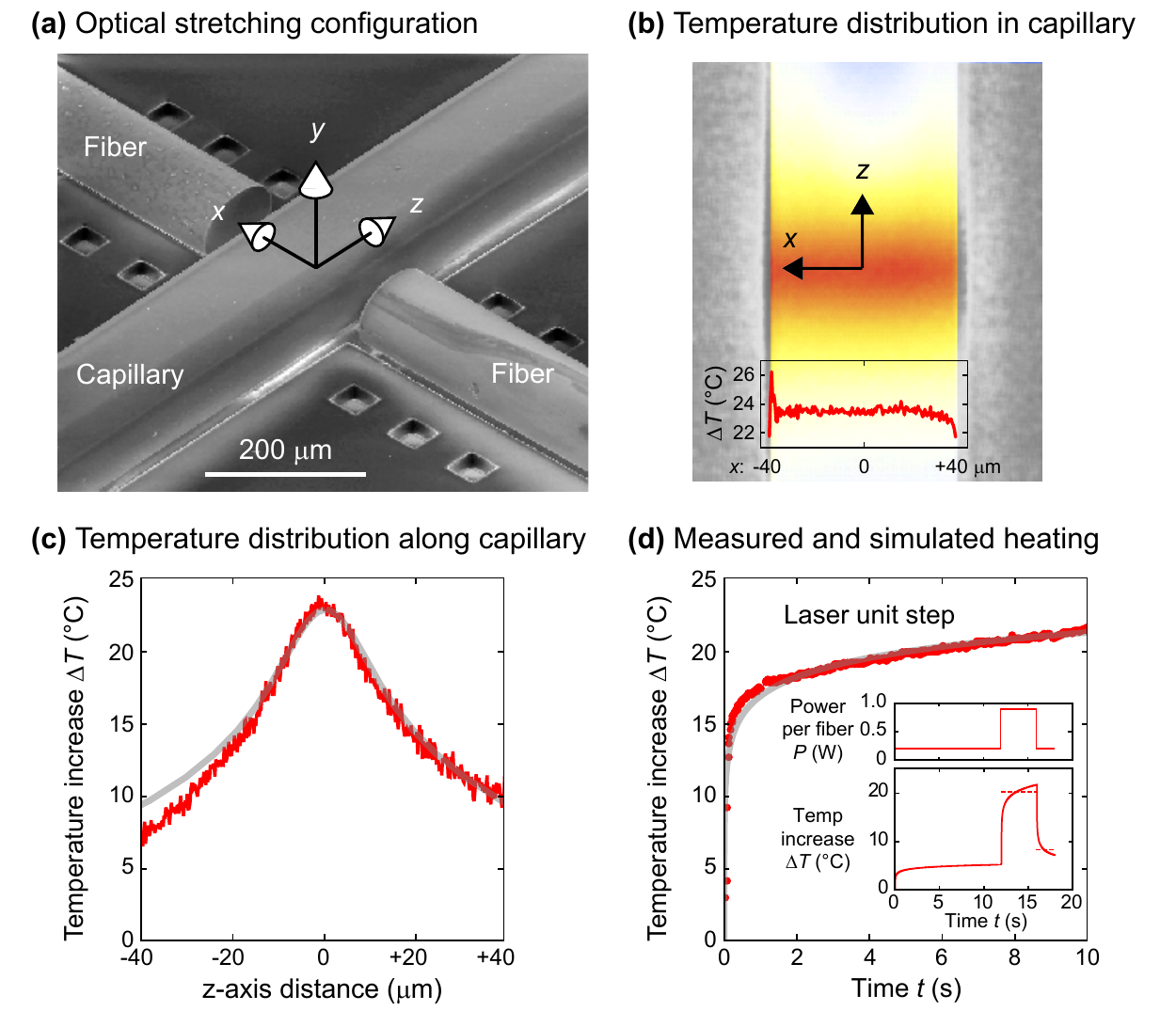}
\caption{(Supplemental) A thermal model of laser-induced cell heating is derived and calibrated by using temperature-sensitive dye. (a)~Coordinate system of laser optical fibers and capillary, where the trapped cell sits at the origin. (b)~Temperature increase from two 1\,W counterpropagating 1064\,nm laser beams, as measured by temperature-sensitive Rhodamine B fluorescent dye. Inset, temperature varies little along the $x$ axis, supporting the use of a relatively simple 1-D axisymmetric model. (c)~Actual and predicted temperature profile along $z$ axis running down the capillary. (Thermal model fit to actual data.) (d)~Actual and predicted temperature increase over time upon laser step increase, supporting the use of the model in predicting average temperatures in response to arbitrary laser profiles over time. Inset, predicted average temperature increase at different periods while stretching a single cell, calculated by convolving laser power profile with impulse response.}
\label{fig:supp-thermal-model}
\end{figure}

Figure~S\ref{fig:supp-thermal-model}(a) reprints Fig.~\ref{fig:OS}(a), 
 now labeled with a coordinate system. The temperature distribution within the capillary can be visualized by plotting the percent attenuation of intensity (with background subtracted) with heat map colorization (Fig.~S\ref{fig:supp-thermal-model}(b)). It can be seen that during laser operation, the spatial temperature distribution at $y=z=0$ along the $x$ axis is minimal; as a result, we consider only spatial changes in the $y-z$ plane, and we further assume axisymmetry (with $r^2=y^2+z^2$) because of the advantage of obtaining an analytically tractable solution. (In reality, due to the geometry of the 80\,$\upmu$m-inside-dimension capillary, at $r>40$\,$\upmu$m the heat dissipation medium is glass for $z=0$, liquid for $y=0$. This difference is ignored.) We will consider the surrounding medium and the cell to have the thermal properties of water; it will emerge that the final general form is minimally sensitive to this assumption.

An energy balance leads to the heat equation
\begin{equation}
\nabla^2 T(r,t)+\frac{Q(r)}{\kappa}=\frac{1}{\alpha}\dot T(r,t)
\label{eq:balance}
\end{equation}
where $T$ is  temperature, $Q$ is the heat generation caused by beam  absorption, $\kappa$ is the thermal conductivity, and $\alpha$ is the thermal diffusivity. (All material properties are assumed to be constant.) 

Volumetric heat generation due to absorption of a Gaussian beam can be modeled as~\cite{peterman2003laser}
\begin{equation}
Q(r)=I(r)A=\frac{4PA}{\pi w^2}e^{-2r^2/w^2}
\end{equation}
where $I(r)$ is the spatial intensity, $A$ is the absorbance at the laser wavelength, $P$ is the power per fiber, and the beam waist or nominal radius $w$ is calculated as described in a previous supplemental section, and is assumed to be constant. (In reality, one beam diverges and the other converges when moving along the $x$ axis.)

It so happens that the spatially varying heat generation term has a relatively simple form in the Hankel domain, generally used to solve integrodifferential equations in cylindrical coordinates, and that the resulting expression for temperature is easily transformed back into the spatial domain. The Hankel transform ($\mathcal{H}[T(r)]\rightarrow T(h)$) of Eq.~\ref{eq:balance} is~\cite{poularikas2009transforms}
\begin{equation}
-k^2 T(h,t)+\frac{PA}{\pi \kappa}e^{-h^2w^2/8}=\frac{1}{\alpha}\dot T(h,t).
\end{equation}
Still in the Hankel domain, the solution for $T(h,t)$ with initial condition $T(h,0)=T_\infty$ can be verified (by using Laplace transforms, for example) to be 
\begin{equation}
T(h,t)=T_\infty+\frac{PA}{\pi\kappa h^2}\left(1-e^{-h^2\alpha t}\right)e^{-h^2w^2/8}.
\end{equation}
At $r=0$, representing the center of the beams and the cell, the inverse Hankel transform is calculated to be~\cite{poularikas2009transforms}
\begin{equation}
T(0,t)=T_\infty+\int_0^\infty h T(h,t)\,dh=T_\infty+\frac{PA}{2\pi\kappa}\ln\left(1+\frac{8\alpha t}{w^2}\right)\approx T_\infty+\frac{PA}{2\pi\kappa}\left[\ln\left(\frac{8\alpha}{w^2}\right)+\ln t\right]
\end{equation}
where this approximation is predicated on the fact that $8\alpha/w^2\gg 1$. We  therefore look for a relationship of the form $T(t)=T_\infty+C_1\ln(1+C_2t)$ with $C_2=8\alpha/w^2\approx 5700\,\mathrm{s}^{-1}$. Figure~S\ref{fig:supp-thermal-model}(c) shows the the numerical spatial solution for general $r$, where $C_1$ has been fit to 2.3$^\circ\mathrm{C}\,(\mathrm{W/fiber})^{-1}$ to match actual data, also shown. The value of $C_1$ is within a factor of 2 of what one would predict by using the properties of water (absorbance $A=15\,\textrm{m}^{-1}$ at 1064\,nm, thermal conductivity $\kappa=0.6\,\textrm{W}\,\textrm{m}^{-1}\,\textrm{K}^{-1}$, thermal diffusivity $\alpha=1.5\times 10^{-7}\,\textrm{m}^2\,\textrm{s}^{-1}$~\cite{hale1973optical}) and the beam Gaussian waist, or nominal radius, $w=w_\mathrm{beam}=14.5\,\upmu$m at $x=0$ as calculated in another supplemental section.

Two simplifying assumptions are noted in the quantitative estimation of cell temperature. First, because the dye fluorescence was measured with an epifluorescence microscope rather than a confocal microscope, brightness is actually integrated through the region $y=[-40,\,40]\,\upmu$m in the capillary. Therefore, attenuation values discussed represent average attenuation over the vertical height. Second, the maximum temperature at $z=0$ is used to characterize laser-induced heating; the average temperature across the whole cell will be lower. Fortunately, errors caused by these two assumptions act in opposite directions and therefore offset each other to some extent, as can be seen in the good agreement between results of laser-induced and external heating in Fig.~\ref{fig:fluidity}(d) 
 and Fig.~S\ref{fig:supp-stiffness-vs-temp}.

Figure~S\ref{fig:supp-thermal-model}(d) shows the good match between the $\ln t$ form discussed above and the measured temperature increase over time for a unit step increase in laser power, supporting the validity of the assumptions made in this derivation. Consequently, we use this fitted model  to estimate the average temperature of the cell in response to arbitrary laser input (Fig.~S\ref{fig:supp-thermal-model}(d, inset)), where we have convolved the impulse temperature response with a more complex laser power profile.

\begin{figure}[h]
\centering
\includegraphics{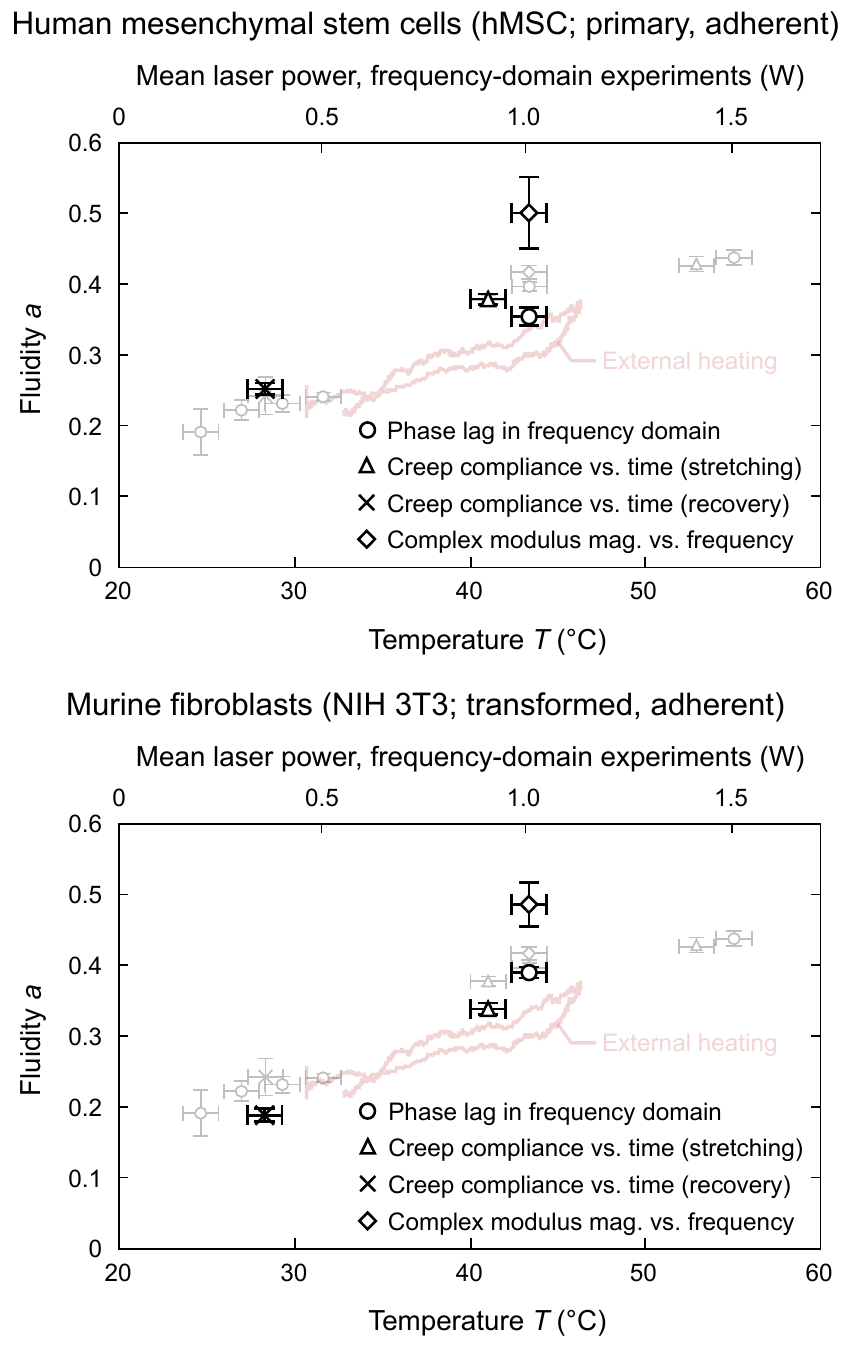}
\caption{(Supplemental, extension of Fig.~\ref{fig:fluidity}.) 
 The finding of increasing fluidity  with increasing temperature, as studied in CH27 lymphoma cells (reprinted here in faint background) is supported in two other eukaryotic cell types via multiple rheological measurements in the time and frequency domains. Fluidity estimates with standard error are tabulated in Table~S1.}
\label{fig:supp-fluidity}
\end{figure}

For an arbitrary power input $P(t)$ per fiber over time, we convolve the power with the temperature vs.\ power  impulse response to obtain $T(t)=T_\infty+\int_0^t P(\tau)\frac{C_1}{1+C_2(t-\tau)}\,d\tau$. As a result, and for the purposes of plotting fluidity vs. temperature in Fig.~\ref{fig:fluidity}(d) 
 and Figs.~S\ref{fig:supp-fluidity} and S\ref{fig:supp-fluidity}, we calculate the average temperature increase $\Delta T=\frac{1}{t_2-t_1}\int_{t_1}^{t_2}[T(t)-T_\infty]\,dt$ (above room temperature  $T_\infty=20\pm 1^\circ$C) to be 20.1$^\circ$C and 32.8$^\circ$C during 0.9\,W and 1.5\,W (per fiber) creep compliance stretching, respectively; 8.5$^\circ$C and 11.2$^\circ$C during 0.9\ W and 1.5\,W (per fiber) creep compliance recovery, respectively; and 23.9$^\circ$C per 1\,W (per fiber) during complex modulus measurements. It is assumed that 10\,s at 0.2\,W per fiber is used to center the cells; this exposure contributes an estimated 5.2$^\circ$C to the increases listed here.

Finally, we address our observation that fluidity decreases at higher load amplitudes, as shown in Fig.~\ref{fig:linearity}(b). 
 The origin of this decrease is not clear, but it could arise if the relationship between fluidity and mean power (corresponding to temperature increase) in Fig.~\ref{fig:linearity}(a). 
  were concave-downward to some extent. We expect  the fluidity value we extract during oscillatory loading to represent the average value; mathematically, the extracted value $a=\int_0^{2\pi} a(P_0+P_1\sin\theta)\,d\theta$ for mean power $P_0$ and amplitude $P_1$. When $a(P)$ is a near-linear function of $P$, the extracted value will be approximately $a(P_0)$; when the function is concave-downward, however, the value will be somewhat less, and larger load amplitudes will produce average fluidity values that are lower than those acquired as smaller amplitudes. Supporting this hypothesis is the finding that fluidity is restored to its small-load-amplitude value when the loading frequency is increased to 10\,Hz to attenuate temperature excursions (Fig.~\ref{fig:linearity}(b)). 

\clearpage
\section*{Supplemental Information: Lumped-component fitting to creep compliance}
\begin{figure}[h]
\centering
\includegraphics{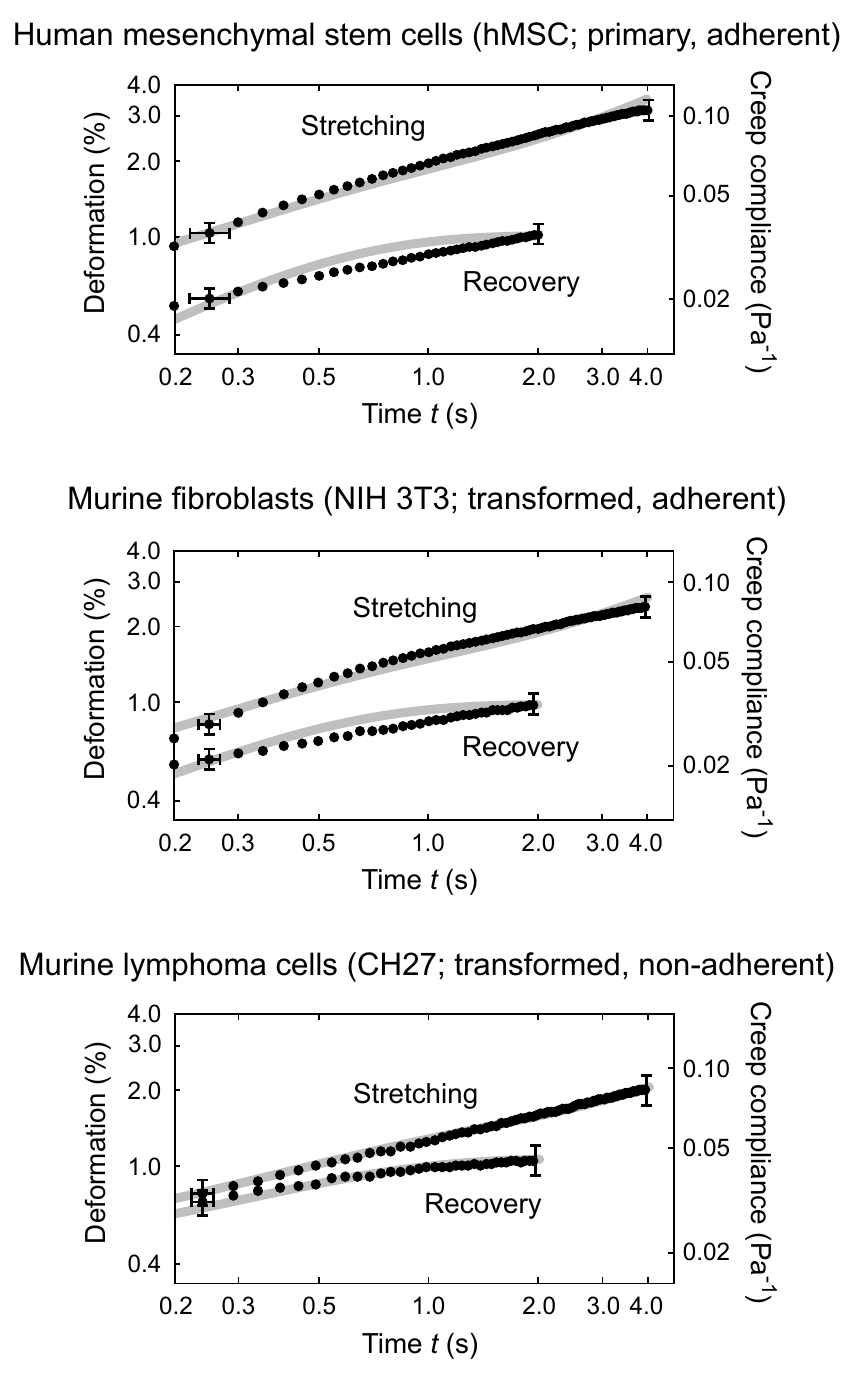}
\caption{(Supplemental, extension of Fig.~\ref{fig:PLR}.) 
 Parameters of a standard linear solid (a series-spring-dashpot pair, $E_1$, $\eta_1$, in series with a parallel-spring-dashpot pair, $E_2$, $\eta_2$) can be fit to the average deformation of single whole cells during and after a unit step in laser-induced stress. A reasonable fit is achieved in the time domain; however, frequency-domain predictions from this parameterization are not borne out (Fig.~\ref{fig:PLR}(b,e,h)), 
  and the extracted time constants $\tau_1$ and $\tau_2$ appear to be artifactual, possibly corresponding to the experiment length rather than any material property. (a) hMSCs: $E_1=69\,\textrm{Pa}$, $\eta_1=55\,\mathrm{Pa\,s}$, $E_2=31\,\mathrm{Pa}$, $\eta_2=11.2\,\mathrm{Pa\,s}$, resulting in time constants $\tau_1=\eta_1/E_1=0.80\,\mathrm{s}$, $\tau_2=\eta_2/E_2=0.36\,\mathrm{s}$. (b) 3T3s: $E_1=77\,\mathrm{Pa}$, $\eta_1=77\,\mathrm{Pa\,s}$, $E_2=33\,\mathrm{Pa}$, $\eta_2=12.8\,\mathrm{Pa\,s}$;  $\tau_1=1.00\,\mathrm{s}$, $\tau_2=0.40\,\mathrm{s}$. (c) CH27s: $E_1=57\,\mathrm{Pa}$, $\eta_1=106\,\mathrm{Pa\,s}$, $E_2=38\,\mathrm{Pa}$, $\eta_2=17\,\mathrm{Pa\,s}$; $\tau_1=1.86\,\mathrm{s}$, $\tau_2=0.45\,\mathrm{s}$.}
\label{fig:spring-dashpot}
\end{figure}

\clearpage
\section*{Supplemental Information: Fluidity vs.\ \emph{in vitro} culture time in primary hMSCs}

\begin{figure}[h]
\centering
\includegraphics{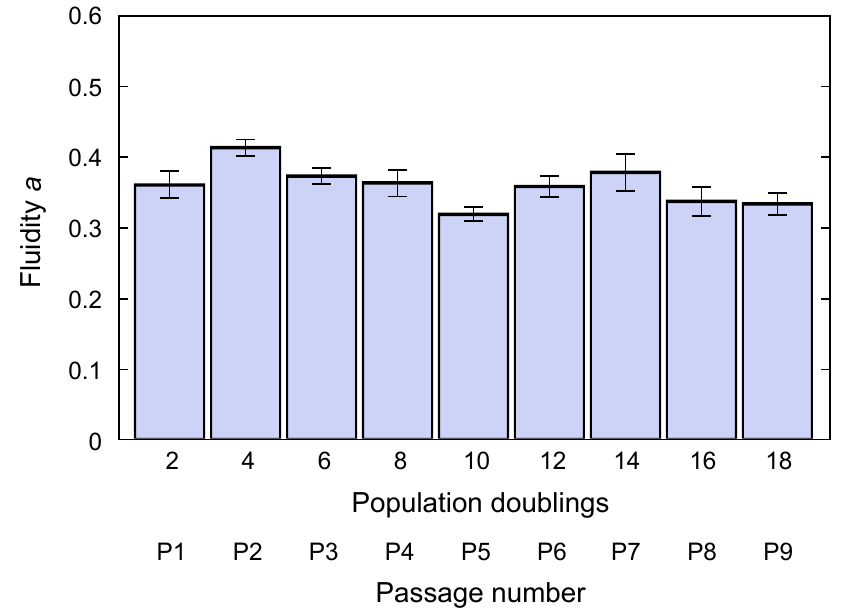}
\caption{(Supplemental) Fluidity is not detectably altered by extended \emph{in vitro} culturing of primary hMSCs.}\label{fig:passaging}
\end{figure}

\clearpage
\section*{Supplemental Information: Storage modulus vs.\ frequency}

\begin{figure}[h]
\centering
\includegraphics{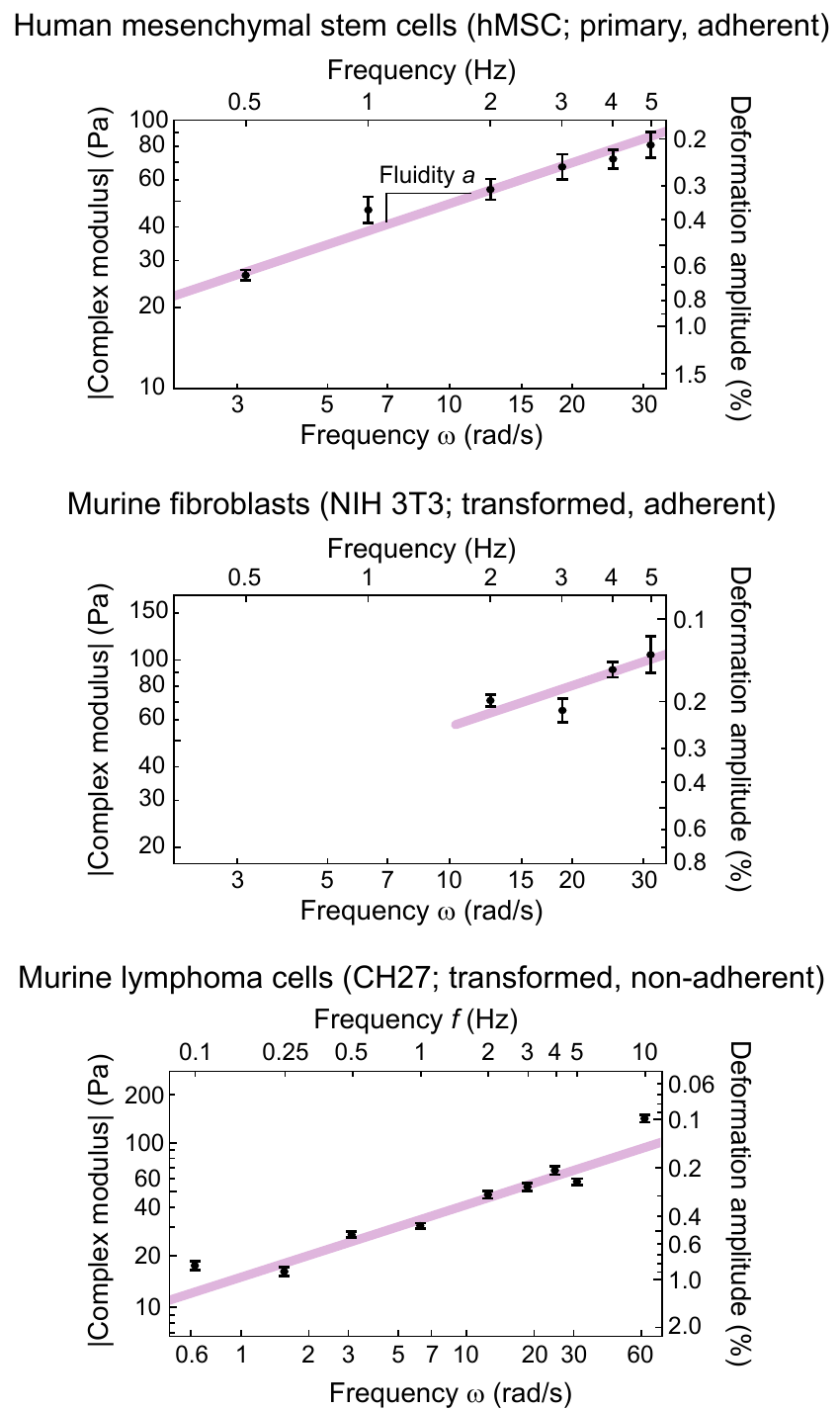}
\caption{(Supplemental) Measurements of frequency-dependent stiffness (acquired at 1\,W per fiber mean, 0.5\,W per fiber sinusoidal amplitude) are compatible with the structural damping (power-law) model of $|G^\star(\omega)|\propto \omega^a$, where $a$ corresponds to the fluidity, for frequencies around 1\,Hz. Rheological model fitting can be challenging when only two (or less than two) decades of time or frequency is available; in this work we leverage  multiple rheological measurements in the time and frequency domains and rank models by consistency across these domains. Fluidity estimates with standard error are tabulated in Table~S1.}\label{fig:storage-modulus}
\end{figure}

\clearpage
\section*{Supplemental Information: Fluidity of individual cells during external heating}
\begin{figure}[h]
\centering
\includegraphics{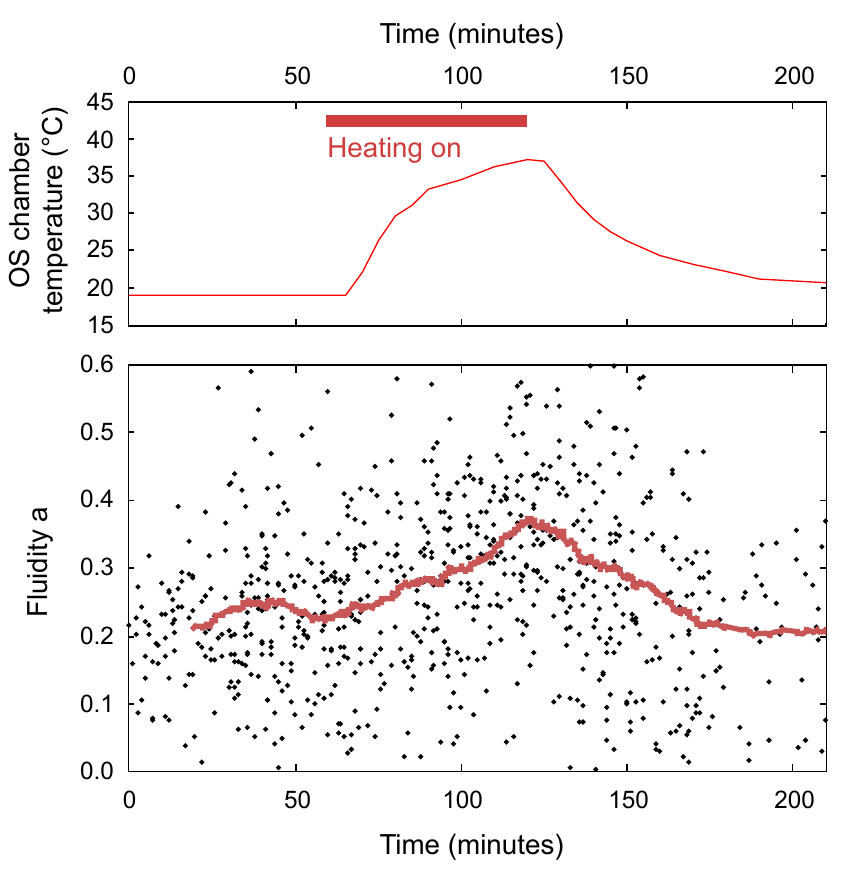}
\caption{(Supplemental, extension of Fig.~\ref{fig:fluidity}.) 
 Chamber temperature, as measured by fluorescent dye, and fluidity values of individual cells ($n=860$), extracted from phase lag, before, during, and after external heating of the optical stretcher chamber. A mean photonic load of 0.5\,W per fiber increased the cell temperature by an additional 11.7$^\circ$C during stretching. Moving average (of 100 cells) indicates trend of increasing fluidity with increasing temperature and vice versa.}
\label{fig:supp-fluidity-vs-time}
\end{figure}

\clearpage
\section*{Supplemental Information: Fluidity vs. temperature for other cell types}

\clearpage
\section*{Supplemental Information: Stiffness vs.\ temperature}
\begin{figure}[h]
\centering
\includegraphics{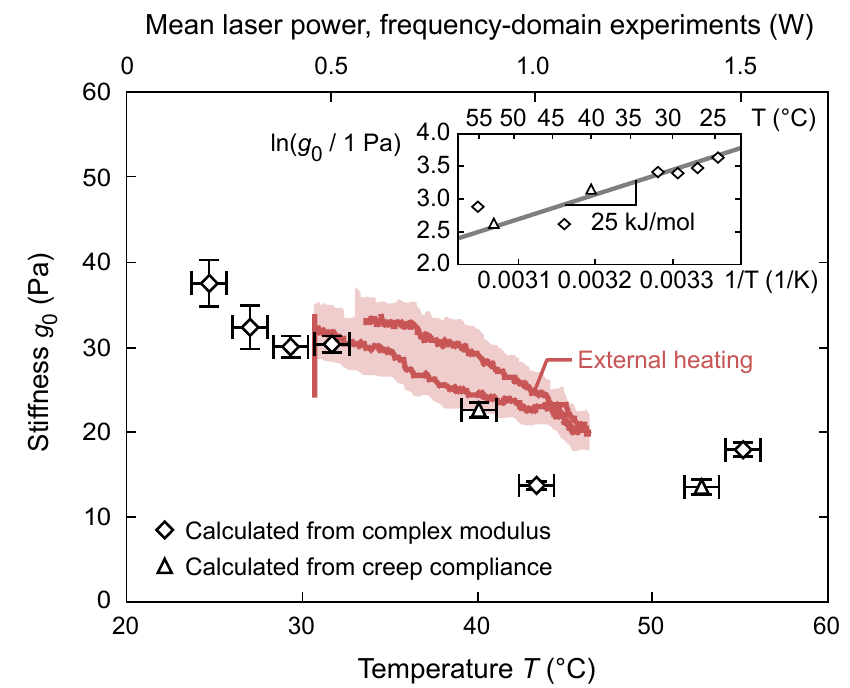}
\caption{(Supplemental, extension of Fig.~\ref{fig:fluidity}.) 
 Measurements in the time and frequency domains under different laser powers, along with experiments performed under external heating, uncoupled to laser power, illustrate the relationship of decreasing stiffness  with increasing temperature  in whole CH27 lymphoma cells. External heating is 100-cell moving average of 860 individual cells with 95\% confidence interval. Inset, determination of activation energy of deformation.}
\label{fig:supp-stiffness-vs-temp}
\end{figure}

The thermorheology of single cells was recently studied under the framework of simple time-temperature superposition (TTS)~\cite{kiessling2013thermorheology}; for example, creep compliance at a non-reference temperature, when plotted against $t/a_T$ rather than $t$ (where $t$ is time and $a_T$ is a scaling constant unrelated to fluidity $a$), was found to overlay creep compliance at a reference temperature.

Simple TTS is not strictly compatible with a temperature-dependent fluidity of a power-law material.  Graphically, simple TTS can be visualized as a translation of the creep vs.\ time relationship on a log-log scale; on this scale, such a translation cannot replicate a line segment with a temperature-dependent slope. However, if fluidity is approximated as a constant over a sufficiently small range, our results are compatible with those acquired under the simple TTS approach, as shown here.

For temperatures $T_i$ around physiological reference temperature 37$^\circ$C, let fluidity $a(T_i)\approx a(T_\textrm{ref})=a$.  Taking creep compliance $J(t,T_i)$ as a thermally activated process with a power-law dependence on time, as shown in Fig.~\ref{fig:PLR}, 
 we have 
\begin{equation}
J(t,T_i)\propto \left(\frac{t}{t_0}\right)^a\exp\left(-\frac{E_J}{RT_i}\right)
\end{equation}
We can obtain $E_J$ (the activation energy for creep compliance) from the slope of $\ln(g_0/1$\,Pa) vs.\ $1/T$, which corresponds to $E_J/R$ under the assumption of constant fluidity. (Complex modulus and creep compliance are inversely related under this assumption.) From Fig.~S\ref{fig:supp-stiffness-vs-temp}(inset) we find that $E_J=25$\,kJ\,mol$^{-1}$.

Simple TTS with time scaling implies that $J(t,T_\mathrm{ref})= J(a_T t,T_i)$, where, for example, $a_T<1$ when deformability and creep compliance are larger at $T_i>T_\mathrm{ref}$~\cite{kiessling2013thermorheology}. Using the constitutive relation for $J(t,T_i)$ above, this expression is equivalent to
\begin{equation}
\left(\frac{t}{t_0}\right)^a\exp\left(-\frac{E_J}{RT_\mathrm{ref}}\right)=\left(\frac{a_Tt}{t_0}\right)^a\exp\left(-\frac{E_J}{RT_i}\right)
\end{equation}
or
\begin{equation}
\ln(a_T)=\frac{E_J}{aR}\left(\frac{1}{T_i}-\frac{1}{T_\mathrm{ref}}\right),
\end{equation}
indicating that the time-scaling parameter $a_T$ should also appear thermally activated with an activation energy of $E_J/a=80$\,kJ\,mol$^{-1}$ for $a\approx 0.3$ at 37$^\circ$C (Fig.~\ref{fig:fluidity}(d)). 
 This is very close to a previous finding of 74\,kJ\,mol$^{-1}$ under the assumption of simple TTS~\cite{kiessling2013thermorheology}, indicating the compatibility of the two approaches at temperatures near $T_\mathrm{ref}=37^\circ$C. We conclude that when the temperature dependence of fluidity is neglected, a material exhibiting power-law rheology is amenable to analysis under the simple TTS framework.

\end{document}